\documentclass[trackchanges,twocolumn,tighten,times,twocolappendix]{aastex701}

\usepackage{mathrsfs}  
\usepackage{amsmath}  
\usepackage{array}
\usepackage{textcomp}

\usepackage[normalem]{ulem}

\usepackage{hyperref}
\newcommand{\hb}{\rm H$\beta$}
\newcommand{\ha}{\rm H$\alpha$}

\newcommand{\feii}{\rm Fe~{\sc ii}}
\newcommand{\oi}{\rm [O~{\sc i}]}
\newcommand{\oiii}{\rm [O~{\sc iii}]}
\newcommand{\nii}{\rm [N~{\sc ii}]}
\newcommand{\sii}{\rm [S~{\sc ii}]}

\newcommand{\rfe}{$\mathcal{R}_{\rm Fe}$}

\newcommand{\mbh}{$M_{\rm BH}$}
\newcommand{\mdot}{$\dot{\mathscr{M}}$}

\def\kms{km\,${\rm s^{-1}}$}

\def\ergscma{\rm erg~s$^{-1}$~cm$^{-2}$~{\AA}$^{-1}$}

\begin{document}

\title{The Population of Type~1 Active Galactic Nuclei in Dark Energy Spectroscopic Instrument.\\
I. Profile-dependent Measurements of the Low-redshift Sample}

\def\authormail{
\href{mailto:wj@nao.cas.cn}{wj@nao.cas.cn}, 
\href{mailto:dwxu@nao.cas.cn}{dwxu@nao.cas.cn}
}

\author[orcid=0009-0000-1228-2373,gname=Zhuheng,sname=Yao]{Zhu-Heng Yao}
\affiliation{National Astronomical Observatories, Chinese Academy of Sciences, 20A Datun Road, Chaoyang District, Beijing 100101, China; \authormail}
\email{zhyao@bao.ac.cn}

\author[orcid=0000-0001-9457-0589,gname=Weijian,sname=Guo]{Wei-Jian Guo}
\affiliation{National Astronomical Observatories, Chinese Academy of Sciences, 20A Datun Road, Chaoyang District, Beijing 100101, China; \authormail}
\email{guowj@bao.ac.cn}

\author[orcid=,gname=Shuo,sname=Zhai]{Shuo Zhai}
\affiliation{National Astronomical Observatories, Chinese Academy of Sciences, 20A Datun Road, Chaoyang District, Beijing 100101, China; \authormail}
\email{zhaishuo@bao.ac.cn}

\author[orcid=0009-0003-1282-9131,gname=Yuchang,sname=Li]{Yu-Chang Li}
\affiliation{School of Physics and Astronomy, Sun Yat-sen University, Zhuhai Campus, 2 Daxue Road, Xiangzhou District, Zhuhai, 519082, China}
\email{liysh96@mail2.sysu.edu.cn}

\author[orcid=0000-0003-0202-0534,gname=Cheng,sname=Cheng]{Cheng Cheng}
\affiliation{National Astronomical Observatories, Chinese Academy of Sciences, 20A Datun Road, Chaoyang District, Beijing 100101, China; \authormail}
\email{chengcheng@bao.ac.cn}

\author[orcid=0000-0003-3086-7804,gname=Junrong,sname=Liu]{Jun-Rong Liu}
\affiliation{Key Laboratory for Particle Astrophysics, Institute of High Energy Physics, Chinese Academy of Sciences, 19B Yuquan Road, Beijing 100049, China}
\affiliation{Dongguan Neutron Science Center, 1 Zhongziyuan Road, Dongguan 523808, China}
\email{}

\author[orcid=,gname=Shuang,sname=Jin]{Shuang Jin}
\affiliation{National Astronomical Observatories, Chinese Academy of Sciences, 20A Datun Road, Chaoyang District, Beijing 100101, China; \authormail}
\email{}

\author[orcid=,gname=,sname=]{Antonios Katsianis}
\affiliation{School of Physics and Astronomy, Sun Yat-sen University, Zhuhai Campus, 2 Daxue Road, Xiangzhou District, Zhuhai, 519082, China}
\email{katsianis@sysu.edu.cn}

\author[orcid=,gname=Jing,sname=Wang]{Jing Wang}
\affiliation{National Astronomical Observatories, Chinese Academy of Sciences, 20A Datun Road, Chaoyang District, Beijing 100101, China; \authormail}
\email{wj@nao.cas.cn}

\author[orcid=,gname=Dawei,sname=Xu]{Da-Wei Xu}
\affiliation{National Astronomical Observatories, Chinese Academy of Sciences, 20A Datun Road, Chaoyang District, Beijing 100101, China; \authormail}
\email{dwxu@nao.cas.cn}

\author[orcid=,gname=Jianyan,sname=Wei]{Jian-Yan Wei}
\affiliation{National Astronomical Observatories, Chinese Academy of Sciences, 20A Datun Road, Chaoyang District, Beijing 100101, China; \authormail}
\email{wjy@nao.cas.cn}


\begin{abstract}

We present a spectral analysis of a large sample of low-redshift type~1 active galactic nuclei (AGNs) selected from the first data release of the Dark Energy Spectroscopic Instrument (DESI~DR1).
For objects at $z\le0.9$, we perform spectral fitting and ML-assisted refinement, yielding 114,754 type~1 AGNs.
We use two fitting schemes that differ only in the broad-line model: the broad Balmer components are fitted with either a Lorentzian function or a sum of Gaussians.
The continuum and {\oiii} luminosity measurements agree closely between the two schemes, whereas the Gaussians scheme generally yields larger broad-{\hb} full width at half maximum (FWHM) values, larger virial black hole masses, and lower accretion rates.
The adopted broad-Balmer profile also changes the decomposition of narrow Balmer emission.
Specifically, the anti-correlation between {\oiii}$\lambda5007$/narrow~{\hb} and accretion rate is stronger under the Gaussians scheme with a Spearman rank correlation coefficient $r_{\rm s}=-0.412$, than under the Lorentzian scheme with $r_{\rm s}=-0.206$; the same tendency is observed in the narrow-line Seyfert~1 galaxies (NLS1s) subset ($r_{\rm s}=-0.435$ and $-0.259$ for the Gaussians and Lorentzian schemes, respectively), suggesting that the Gaussians scheme may provide a more self-consistent separation of the broad and narrow {\hb} components.
Moreover, the consistent correlations recovered under both profile prescriptions favor a continuous type~1 accretion-state sequence, with NLS1s occupying its high-accretion end rather than constituting a distinct population.

\end{abstract}




\section{Introduction}
\label{sec:intro}
Active galactic nuclei (AGNs) provide a unique laboratory for studying black-hole growth, accretion physics, and the co-evolution of supermassive black holes (SMBHs) and their host galaxies~\citep{Ho2008,WangWei2008,AlexanderHickox2012,Fabian2012,KormendyHo2013,HeckmanBest2014}.
In the optical spectra of AGNs, narrow forbidden lines are observed in both type~1 and type~2 sources and trace the larger-scale narrow-line region (NLR;~\citealt{Koski1978,Schmitt2003,Groves2004,BaskinLaor2005,Wang2011,Netzer2015,Zhu2024}), whereas type~1 AGNs additionally show broad permitted lines from the broad-line region (BLR).
In particular, broad Balmer lines and optical {\feii} emission probe the low-ionization BLR, providing diagnostics of the gas geometry and kinematics close to the central black hole~\citep{KwanKrolik1981,BorosonGreen1992,Marziani2010}.
The full width at half maximum (FWHM) of broad {\hb} is widely used to characterize the line-of-sight velocity of the {\hb}-emitting BLR gas and to estimate black-hole masses~\citep[e.g.,][]{BlandfordMcKee1982,Peterson2004,Grier2017,Bao2022,Li2024,Hu2025}, especially for single-epoch virial black-hole mass estimators~\citep[e.g.,][]{GreeneHo2005,VestergaardPeterson2006,McGill2008,Park2012,Bentz2013,DallaBonta2020}.
$\mathcal{R}_{\rm Fe}=F_{\rm Fe~{\textsc{ii}}}/F_{\rm broad\,H\beta}$ is often used as an empirical tracer of accretion state and has been incorporated into accretion-rate-dependent BLR radius calibrations~\citep{DuWang2019}.
Meanwhile, these two quantities define a key observational plane in the Eigenvector~1 (EV1), or quasar main-sequence~\citep{BorosonGreen1992,Sulentic2002,Marziani2010,Zamfir2010,ShenHo2014,Panda2019}.
Therefore, accurate measurements of broad-{\hb} FWHM and {\rfe} are essential for inferring black-hole mass and accretion-related quantities.

The choice of broad-line profile plays an important role in these fitting measurements.
Previous studies have shown that broad-line shape varies across the EV1 sequence:
Population~A sources, particularly those at the low-FWHM end, are frequently well described by Lorentzian-like profiles, whereas Population~B sources often require multiple broad components, including a very broad component~\citep{Sulentic2002,Marziani2010,Zamfir2010}.
These profile differences may reflect the kinematics and viewing geometry of the BLR.
\citet{KollatschnyZetzl2013} modeled broad-line shapes using intrinsic Lorentzian profiles associated with turbulent motions and further broadened by rotation.
In a bowl-shaped BLR model, \citet{Goad2012} showed that turbulence can produce Lorentzian-like profiles at low inclinations, while increasing inclination enhances the contribution of projected rotational motions and produces broader, more Gaussian-like profiles.
Thus, different profile prescriptions can assign different widths to the same observed profile, affecting the measured and derived quantities such as broad-{\hb} FWHM, {\rfe}, virial black-hole mass, and accretion-rate estimates.
Previous comparisons of Lorentzian and two-Gaussian measurements in narrow-line Seyfert~1 galaxies (NLS1s) have revealed systematic differences in broad-{\hb} FWHM and the inferred virial black-hole masses~\citep{Bian2008}.
Moreover, because narrow {\hb} is superposed on the broad-{\hb} base, the adopted broad-line profile can redistribute flux between the broad and narrow components and consequently alter narrow-line diagnostic ratios such as {\oiii}$\lambda5007$/narrow {\hb}.
A similar decomposition issue arises in the {\ha} region, where narrow {\ha} is embedded within the broad {\ha}+{\nii} complex, introducing uncertainty into the {\ha}-based Baldwin--Phillips--Terlevich (BPT) diagnostic ratios~\citep{BPT1981}.

This profile dependence is especially relevant for NLS1s, whose definition is tied directly to the measured broad-{\hb} width.
NLS1s are conventionally identified by ${\rm FWHM}_{\rm broad\,H\beta}\lesssim2000$~{\kms}~\citep{Goodrich1989} and commonly exhibit strong optical {\feii} emission and relatively weak {\oiii} emission~\citep{OsterbrockPogge1985,BorosonGreen1992,Grupe2000,Xu2012}.
However, the conventional 2000~{\kms} FWHM threshold is an empirical criterion, and objects close to this boundary may be sensitive to the adopted broad-line fitting prescription~\citep{MullaneyWard2008}.
Although large Sloan Digital Sky Survey (SDSS) NLS1 catalogs~\citep{Zhou2006,Rakshit2017,Paliya2024} have enabled statistical studies of NLS1 properties, the effects of broad-line profile prescriptions have received limited systematic consideration in these samples.
This motivates us to examine how broad-line profile prescriptions affect the statistical measurements and classifications of type~1 AGNs, and whether NLS1s can be regarded as a physically distinct population.

The Dark Energy Spectroscopic Instrument (DESI) provides a powerful data set for addressing these questions.
DESI is a highly multiplexed spectroscopic survey instrument designed to obtain optical spectra for millions of galaxies and quasars over a wide wavelength range~\citep{DESI2016,DESI2022_overview}.
Its first Data Release (DR1) contains high-confidence redshifts for 18.7 million objects, including approximately 1.6 million spectroscopically classified quasars, and provides uniformly reduced, flux-calibrated spectra and pipeline redshift classifications~\citep{DESI2024_EDR,DESI2025_DR1}.
These features make it possible to fit the same objects with different broad-line profile prescriptions and to quantify their effects on emission-line measurements, virial quantities, and narrow-line flux ratios in a statistical manner.

In this work, we construct a low-redshift broad-{\hb} type~1 AGN sample from DESI~DR1 spectra, which is further refined using a machine-learning-assisted (ML-assisted) procedure.
Each spectrum is decomposed in the {\hb}--{\oiii} region, and the {\ha}--{\nii} region is also fitted for the lower-redshift subset.
To assess the effects of the broad-Balmer profile adopted in the spectral fitting, we employ two schemes in which the broad Balmer lines are fitted with either a Lorentzian profile or a sum of two Gaussian components (hereafter the Lorentzian and Gaussians schemes), while the continuum, {\feii} emission, narrow-line components, and their associated fitting constraints are treated identically.
We then compare the resulting distributions and correlations of the measured emission-line properties and derived physical quantities between the two fitting schemes.

The paper is organized as follows.
Section~\ref{sec:sample} describes the construction of the type~1 sample, including the DESI parent-sample selection, spectral-fitting procedure, ML-assisted refinement, and final sample definition.
Section~\ref{sec:measurements} presents the main emission-line measurements and derived physical quantities, including the broad- and narrow-line properties, optical diagnostic diagrams, line-luminosity scaling relations, black-hole masses and accretion rates.
Section~\ref{sec:physical_implications} discusses the effects of the adopted broad-Balmer profile on the {\rfe}–FWHM plane, NLS1 classification, virial quantities, equivalent-width correlations, and narrow-line flux-ratio trends.
The main conclusions are summarized in Section~\ref{sec:sum}.
Throughout this work, we assume a cosmology with $H_0=70~{\rm km~s^{-1}~Mpc^{-1}}$ and $\Omega_{\rm M}=0.3$.

\section{Sample Construction}
\label{sec:sample}

\subsection{Parent Sample}
\label{subsec:parent_sample}

\begin{figure*}[htbp]
\centering
\includegraphics[width=\textwidth, angle=0]{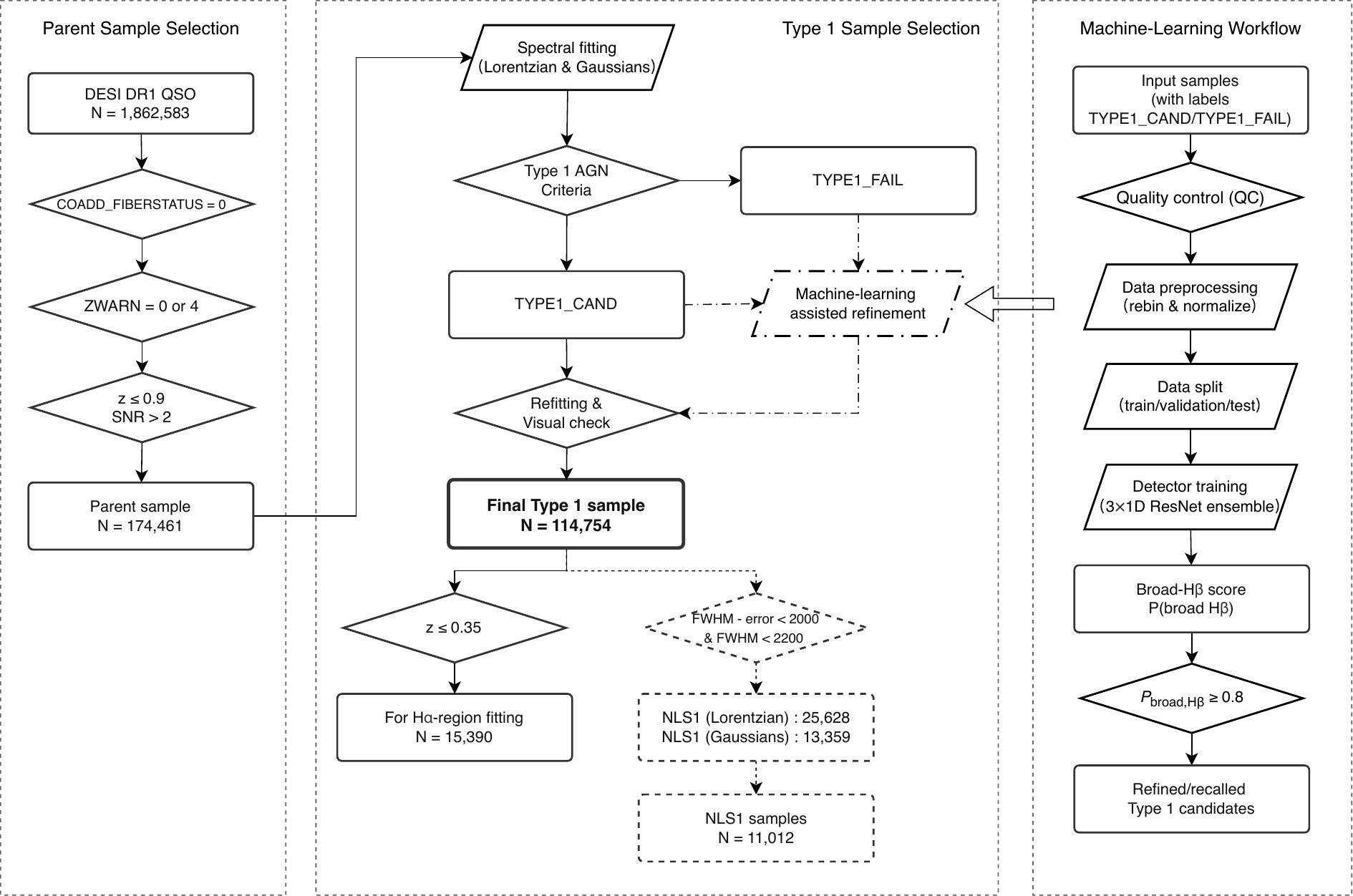}
\caption{
Flowchart of the type~1 AGN sample construction procedure.
The three dashed boxes, from left to right, show the parent sample selection, type~1 sample selection, and the ML-assisted refinement workflow.
Starting from the DESI~DR1 \texttt{QSO}-classified spectra, we apply the basic data-quality, redshift, and SNR criteria, including \texttt{COADD\_FIBERSTATUS}$=0$, \texttt{ZWARN}$=0$ or $4$, $z\le0.9$, and $\rm SNR>2$.
The {\hb}–{\oiii} region is then decomposed using the Lorentzian and Gaussians fitting schemes to identify the initial fitting-based type~1 candidates.
The ML score is used to flag low-confidence candidates and possible recovered type~1 objects for re-evaluation.
Within the fitting-based selection box, the dashed branch indicates the profile-dependent NLS1 subdivision, which are used only in the subsequent population analysis.
}
\label{fig:selproc_desitype1}
\end{figure*}

Figure~\ref{fig:selproc_desitype1} summarizes the full sample-construction procedure, from the DESI~DR1 spectra to the final type~1 AGN sample.
We first constructed the parent sample from the full set of 28,425,963 spectra in DESI~DR1.
Spectra classified as \texttt{QSO} by the DESI pipeline were selected as the starting sample, in order to focus on objects most likely to show AGN signatures.
We then required \texttt{COADD\_FIBERSTATUS}$=0$ to exclude spectra with known fiber- or coaddition-quality problems, and adopted \texttt{ZWARN}$=0$ or $4$ to retain spectra with reliable pipeline redshifts while allowing the accepted QSO-related warning flag.
After these quality and redshift-flag cuts, the sample contains 1,751,038 QSO spectra.
To ensure that the broad {\hb} and {\oiii} emission lines, together with the surrounding fitting window, remain within the DESI wavelength coverage, we further restricted the sample to $z\le0.9$, leaving 301,531 spectra.
Each spectrum was corrected for Galactic extinction using the dust map of~\citet{SchlaflyFinkbeiner2011} and the extinction law of~\citet{Cardelli1989}, adopting $R_V=3.1$.
The extinction-corrected spectra were then shifted to the rest frame using the adopted DESI redshifts.
Finally, we measured the spectral signal-to-noise ratio (SNR) around rest-frame 5100~{\AA} directly from the corrected spectra and required $\mathrm{SNR}>2$, removing spectra whose continuum quality is too low for reliable spectral decomposition.
The resulting parent sample contains 174,461 objects and forms the input sample for the subsequent spectral decomposition.

\subsection{Spectral Fitting Scheme}
\label{subsec:spec_fit}

We adopted a spectral-decomposition and emission-line fitting procedure to separate the host-galaxy, AGN continuum, and emission-line components.
First, the observed spectrum was modeled as the sum of a host-galaxy stellar component and an AGN continuum, and the best-fitting host contribution was subtracted.
Second, the host-subtracted spectrum was fitted in the {\hb}–{\oiii} region to measure the local AGN continuum, optical {\feii} emission, and broad- and narrow-line components.
For the lower-redshift subset with {\ha} coverage, the {\ha}–{\nii} region was also fitted.
This procedure reduces contamination from host-galaxy starlight while preserving the AGN continuum and emission-line components for the subsequent analysis.

\subsubsection{Host-galaxy Decomposition}
\label{subsubsec:fithost}

At low redshift, the optical spectra of AGNs can contain a substantial contribution from host-galaxy starlight.
We therefore modeled and subtracted the host-galaxy component before performing the detailed AGN emission-line fitting~\citep{Rakshit2017}.
Within continuum-dominated wavelength intervals, with strong emission-line regions masked, the observed spectrum was modeled as the sum of a host-galaxy stellar component and an AGN featureless continuum,
\begin{equation}
F_{\lambda}^{\rm obs} = F_{\lambda}^{\rm host} + F_{\lambda}^{\rm AGN}.
\end{equation}
The host-galaxy component was represented by a linear combination of simple stellar population (SSP) templates,
\begin{equation}
F_{\lambda}^{\rm host} = \sum_i a_i T_{i,\lambda},
\end{equation}
where $a_i$ is the coefficient of the $i$th template and $T_{i,\lambda}$ denotes the corresponding SSP spectrum resampled onto the wavelength grid of the observed spectrum.
The AGN continuum was approximated by a power law,
\begin{equation}
F_{\lambda}^{\mathrm{AGN}} = A_\mathrm{PL} \left( \frac{\lambda}{5100\,\mathrm{\AA}} \right)^{\alpha},
\end{equation}
where $A_{\mathrm{PL}}$ is the normalization and $\alpha$ is the spectral index.
The SSP templates were taken from~\citet{BruzualCharlot2003}.
Our template library contains 30 SSP spectra constructed from ten stellar-population ages (5 Myr, 25 Myr, 100 Myr, 290 Myr, 640 Myr, 900 Myr, 1.4 Gyr, 2.5 Gyr, 5 Gyr, and 11 Gyr) and three metallicities ($Z=0.008$, $0.02$, and $0.05$).
For each object, the templates were interpolated onto the wavelength grid of the observed spectrum and normalized around 5100~{\AA}.
The SSP coefficients were constrained to be non-negative, so that the host contribution was represented by a non-negative linear combination of the 30 templates.
The model with the minimum reduced $\chi^2$ was adopted as the best-fitting solution.

\begin{figure}
\centering
\includegraphics[width=\columnwidth, angle=0]{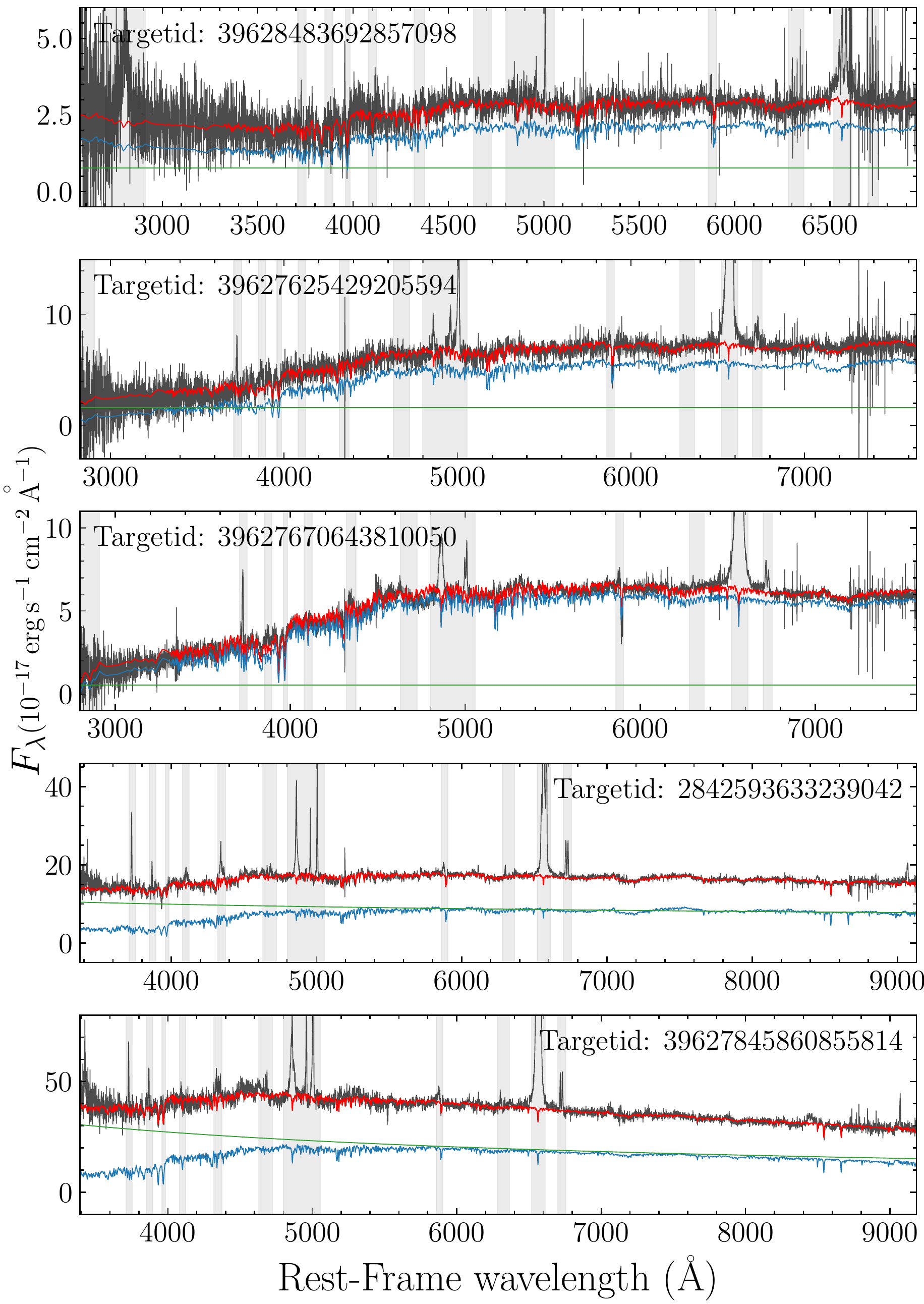}
\caption{
Representative examples of the host-galaxy decomposition for five objects in the parent sample.
In each panel, the black curve shows the extinction-corrected rest-frame spectrum, the red curve shows the best-fitting total continuum model, the blue curve represents the host-galaxy stellar component, and the green curve represents the AGN power-law continuum.
The gray shaded regions mark the wavelength intervals masked during the decomposition because of strong emission lines.
}
\label{fig:fithost}
\end{figure}

Representative decompositions are shown in Figure~\ref{fig:fithost}.
We emphasize that this procedure is intended to provide a practical and homogeneous correction for host-galaxy starlight contamination in a large spectroscopic sample, rather than a detailed reconstruction of the stellar-population parameters or stellar kinematics for individual objects.

\subsubsection{AGN Spectral Fitting}
\label{subsubsec:fitagn}

After subtracting the host-galaxy contribution, we fitted the host-subtracted spectra in the {\hb}--{\oiii} region over the rest-frame wavelength range 4430--5550~{\AA}.
The fitting was performed with \texttt{DASpec}\footnote{\texttt{DASpec} is available at:\\ \url{https://github.com/PuDu-Astro/DASpec}}~\citep{Du2024_daspec}, which performs multicomponent spectral fitting through $\chi^2$ minimization using the Levenberg--Marquardt algorithm.
To quantify the dependence of the measured quantities on the adopted broad-line model, each spectrum was fitted independently using two prescriptions for the broad {\hb} component:
a single Lorentzian profile and a multi-Gaussian profile represented by the sum of two Gaussians.
For brevity, these two prescriptions are referred to as the Lorentzian and Gaussians fitting schemes throughout this paper.
All other spectral components and fitting constraints were kept identical between the two fitting schemes.

The complete fitting model consists of:
(1) a power-law component representing the AGN continuum;
(2) an optical {\feii} template from~\citet{BorosonGreen1992}, broadened by convolution with a Gaussian kernel;
(3) a broad {\hb} component modeled either by Lorentzian or Gaussians;
(4) a single Gaussian component for narrow {\hb};
(5) single Gaussian components for {\oiii}$\lambda4959$ and $\lambda5007$.
The two {\oiii} components were required to share the same velocity shift and line width, and their flux ratio {\oiii}$\lambda4959$-to-$\lambda5007$  was fixed to the theoretical value of 1:3.
The narrow {\hb} component was tied to the {\oiii} component, such that they shared the same velocity shift and line width.
The broad {\hb} component was allowed to vary independently of the narrow-line components.
We note that a single Gaussian may not fully capture asymmetric wings or multiple kinematic components in the {\oiii} profiles, and this limitation may affect the broad {\hb} measurements.
Nevertheless, we adopt this simplified model to provide a uniform treatment of narrow lines in both fitting schemes, and a detailed study of the {\oiii} profiles and narrow-line kinematics is planned for future work.

\begin{figure*}[!ht]
\centering
\includegraphics[width=\textwidth, angle=0]{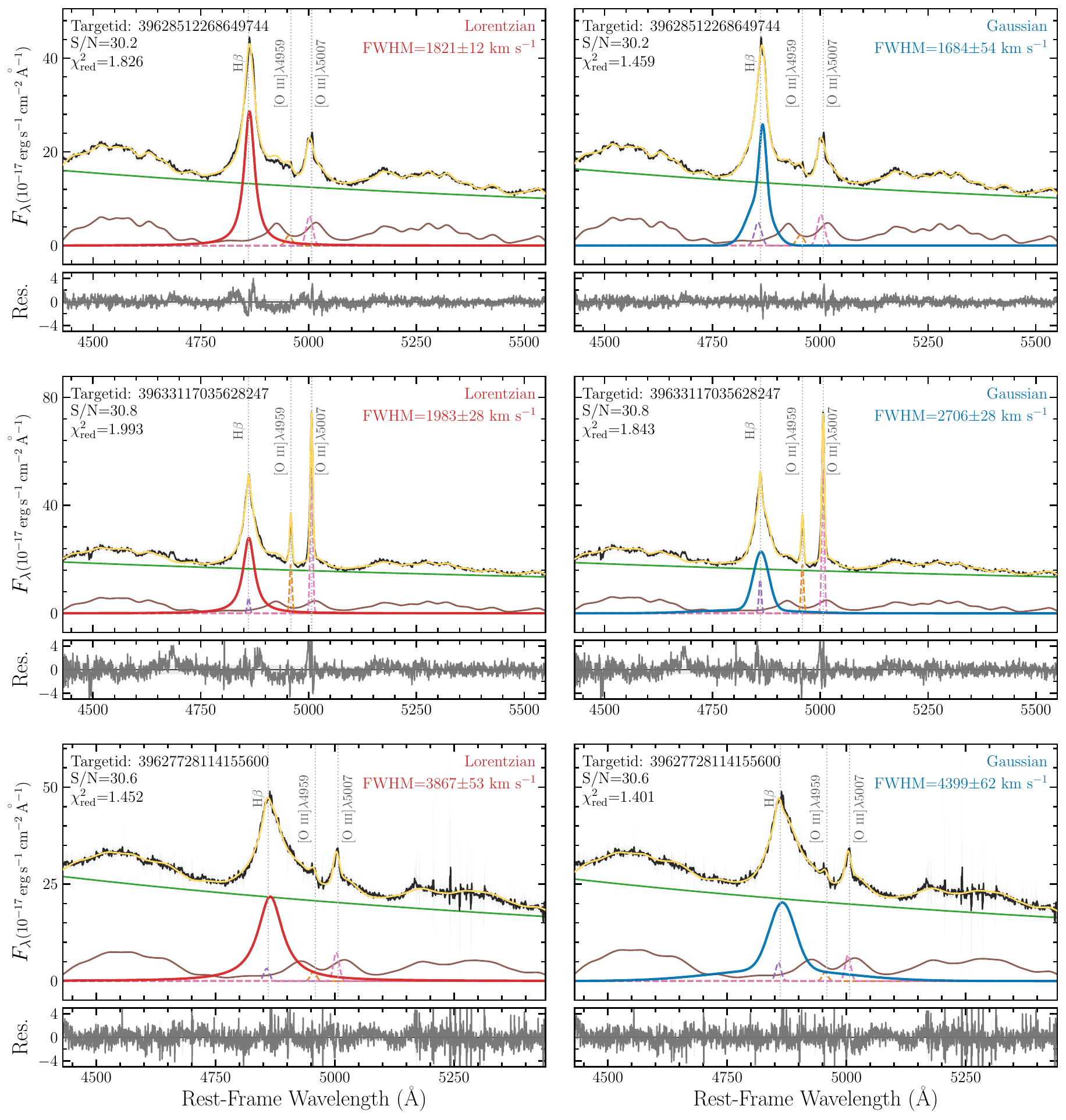}
\caption{
Examples of {\hb}--{\oiii} spectral decompositions for three type~1 AGNs.
For each object, the left and right panels show the Lorentzian and Gaussians schemes, respectively.
The black curve is the host-subtracted rest-frame spectrum, the yellow curve is the total best-fitting model, the green curve represents the power-law continuum, and the brown curve shows the optical {\feii} template.
The red and blue solid curves denote the broad-{\hb} components in the Lorentzian and Gaussians fits, respectively.
The narrow {\hb} and {\oiii} components are shown by dashed curves, and the vertical dotted lines mark {\hb}, {\oiii}$\lambda4959$, and {\oiii}$\lambda5007$.
The lower subpanels show the residuals.
The SNR, reduced $\chi^2$ evaluated within the fitting window and measured broad-{\hb} FWHM are indicated in each panel.}
\label{fig:fitexamples}
\end{figure*}

For the Gaussians fitting scheme, the reported broad-{\hb} flux and FWHM were measured from the combined profile formed by the sum of the two Gaussian components, rather than from either individual Gaussian.
Because the continuum, {\feii}, narrow-line models, wavelength masks, and fitting constraints were otherwise identical, the differences between the two sets of measurements primarily reflect the adopted broad-{\hb} profile prescription.
Representative examples of the Lorentzian and Gaussians schemes are presented in Figure~\ref{fig:fitexamples}, and reduced $\chi^2$ were calculated within the corresponding fitting windows.

\begin{figure*}[!ht]
\centering
\includegraphics[width=\textwidth, angle=0]{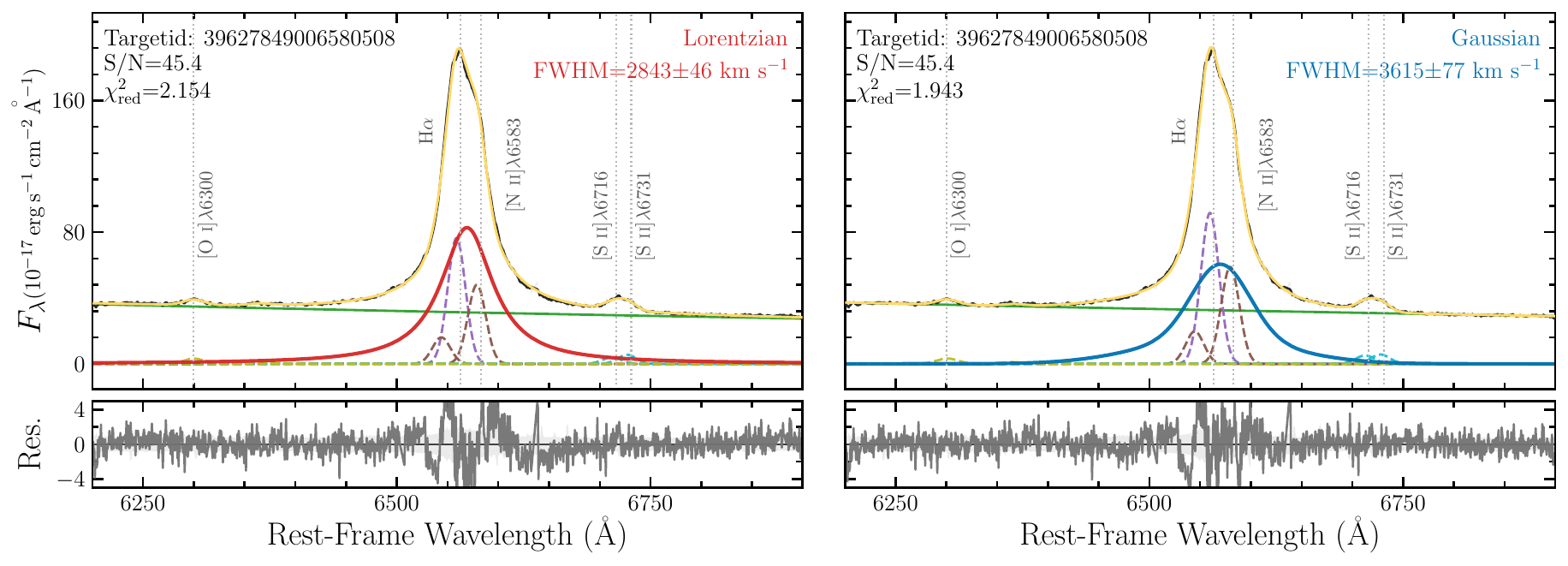}
\caption{
Examples of {\ha}-region spectral decompositions.
The plotting conventions are the same as in Figure~\ref{fig:fitexamples}, but for the {\ha} fitting window.
The vertical dotted lines mark {\oi}$\lambda6300$, {\ha}, {\nii}$\lambda6583$, and {\sii}$\lambda\lambda6716,6731$.
The SNR, reduced $\chi^2$, and measured broad-{\ha} FWHM are labeled in each panel.
}
\label{fig:fitexample_ha}
\end{figure*}

For the lower-redshift subset in which the {\ha} region is covered, we additionally fitted the rest-frame 6200--6900~{\AA} region.
The model includes a local power-law continuum, a broad {\ha} component modeled with the same broad-line prescription as broad {\hb} (i.e. the Lorentzian and Gaussians schemes), a narrow {\ha} component, the {\oi} $\lambda\lambda6300,6364$ doublet, the {\nii} $\lambda\lambda6548,6583$ doublet, and the {\sii} $\lambda\lambda6716,6731$ doublet.
The narrow {\ha}, {\oi}, {\nii}, and {\sii} components were modeled with single Gaussians.
The {\oi} $\lambda6364$-to-$\lambda6300$ and {\nii} $\lambda6548$-to-$\lambda6583$ flux ratios were fixed to the theoretical value of 1:3, while the {\sii} doublet flux ratio was left free.
The narrow {\ha} and forbidden narrow-line components were tied to share the same velocity shift and line width, whereas the broad {\ha} component was allowed to vary independently.
Representative {\ha}-region decompositions are shown in Figure~\ref{fig:fitexample_ha}.

\subsection{Type~1 Sample}
\label{subsec:type1_sample}

We required a successful decomposition of this spectral window and a reliable broad-{\hb} detection, defined by a positive integrated broad-{\hb} flux, $F_{\rm broad\,H\beta}/\sigma_F \ge 3$, and ${\rm FWHM}_{\rm broad\,H\beta}/{\rm FWHM\ Err}_{\rm broad\,H\beta} \ge 3$.
Objects satisfying these criteria were initially labeled as \texttt{TYPE1\_CAND}, whereas the remaining objects were provisionally labeled as \texttt{TYPE1\_FAIL}.

Based on these fitting-based labels, we applied an ML-assisted refinement procedure to improve the reliability and completeness of the type~1 selection.
The ML score was used as a supplementary indicator of broad-{\hb} spectral morphology, rather than as a standalone source classifier or a replacement for spectral fitting;
its construction and performance are described in Appendix~\ref{app:ml}.
Low-scoring \texttt{TYPE1\_CAND} objects and high-scoring \texttt{TYPE1\_FAIL} objects were flagged for re-evaluation.
The flagged spectra were refitted and visually inspected, and only objects with convincing broad-{\hb} features satisfying the fitting-quality requirements were retained in or added to the type~1 sample.

After the fitting-based selection, ML-assisted refinement, refitting, and inspection of the flagged spectra, the final working sample contains 114,754 type~1 AGNs.
The corresponding machine-readable catalog provides basic object information and two parallel sets of Lorentzian- and Gaussians-based spectral measurements;
the full column descriptions are given in Appendix~\ref{app:catalog_columns}.

Figure~\ref{fig:dist-zsnrdn4000} compares the redshift, SNR, and raw $D_n(4000)$ distributions of the final type~1 sample with those of the 174,461-object parent sample.
Here, $D_n(4000)$ is the narrow 4000~{\AA} break index, defined as the ratio of the mean flux densities in the 4000--4100~{\AA} and 3850--3950~{\AA} bands~\citep{Balogh1999}.
The selected type~1 AGNs span nearly the full redshift and SNR ranges of the parent sample, but their relative fraction decreases toward the highest redshifts, which may be related to that the {\hb}--{\oiii} fitting window approaches the red end of the DESI spectral coverage and broad-{\hb} detection becomes less stable.
In addition, compared with the parent sample, the type~1 sample is more concentrated at smaller raw $D_n(4000)$ values, consistent with a stronger contribution from the featureless AGN continuum and/or weaker host-galaxy dilution.

\begin{figure*}
\centering
\includegraphics[width=\textwidth, angle=0]{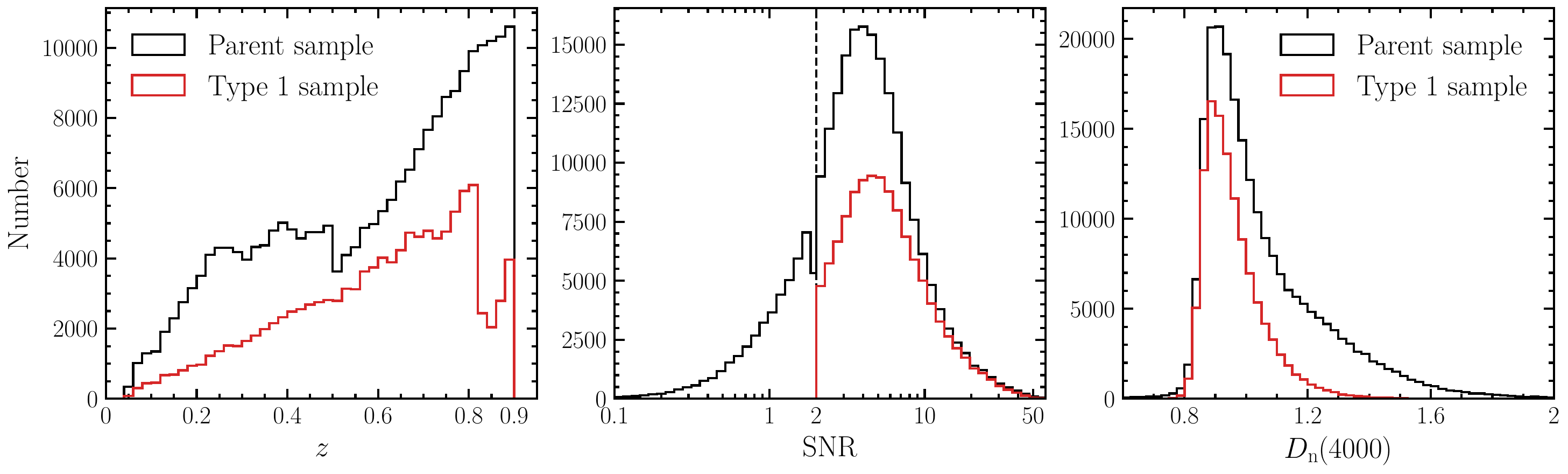}
\caption{
Distributions of redshift (left), SNR (middle), and raw $D_n(4000)$ (right) for the 174,461-object parent sample and the final type~1 sample.
The black and red histograms represent the parent and type~1 samples, respectively.
The vertical dashed line in the middle panel marks the adopted threshold of $\mathrm{SNR}=2$.
}
\label{fig:dist-zsnrdn4000}
\end{figure*}

Among the final type~1 sample, 15,390 objects at $z\le0.35$ are used for the {\ha}-region analysis.
We also define a conservative profile-stable NLS1 subset by taking the intersection of the objects classified as NLS1s under both fitting schemes;
this subset contains 11,012 objects and is used as the NLS1 comparison sample throughout the analysis (see details in Section~\ref{subsec:nls1_class}).

\section{Emission-line Measurements and Physical Properties}
\label{sec:measurements}

\begin{figure*}
\centering
\includegraphics[width=\textwidth, angle=0]{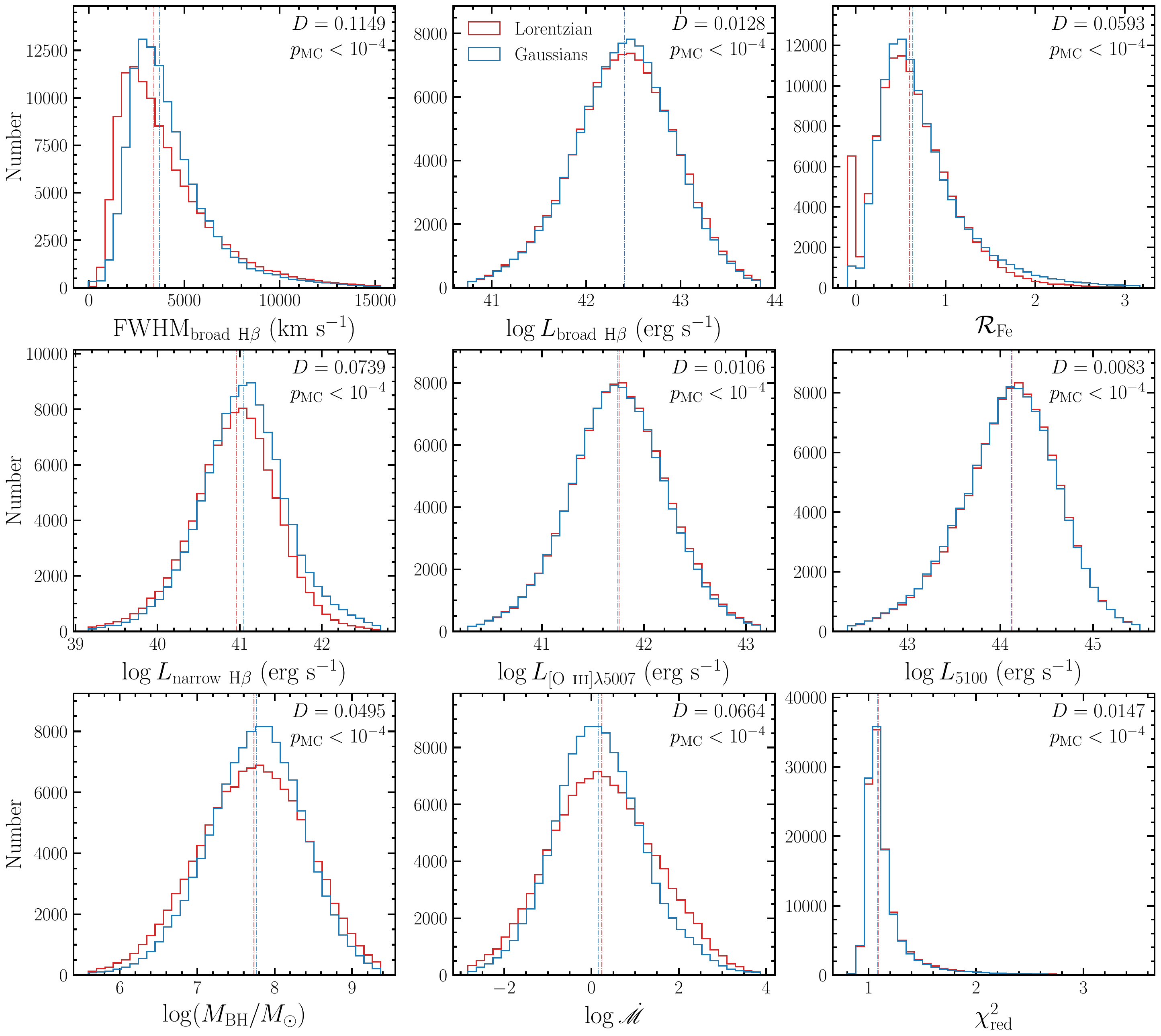}
\caption{
Distributions of the main emission-line measurements and derived physical quantities for the DESI low-redshift broad-{\hb} type~1 AGN sample.
The red and blue histograms show the results obtained from the Lorentzian and Gaussians broad-{\hb} fitting schemes, respectively.
The panels show, from left to right and from top to bottom, the broad-{\hb} FWHM, broad-{\hb} luminosity, {\feii} strength {\rfe}, narrow {\hb} luminosity, {\oiii} $\lambda5007$ luminosity, host-subtracted 5100~{\AA} continuum luminosity, virial black-hole mass, dimensionless accretion rate and reduced $\chi^2$.
The vertical dashed lines mark the median values of the corresponding distributions.
The annotations show the KS statistic $D$ and its Monte Carlo $p$-value, $p_{\rm MC}$, calibrated using 10,000 paired bootstrap resamples.
}
\label{fig:dist-fitl2dg}
\end{figure*}

\subsection{Broad Emission Lines}
\label{subsec:bel}

The broad Balmer emission and optical {\feii} emission provide the key diagnostics of the low-ionization BLR in our DESI broad-{\hb} type~1 AGN sample. 
Figure~\ref{fig:dist-fitl2dg} summarizes the distributions of the main broad-line quantities measured from the Lorentzian and Gaussians fitting schemes, and presents the results of Kolmogorov--Smirnov (KS) tests assessing whether the distributions of each parameter differ significantly between the two fitting schemes.
The broad-{\hb} FWHM spans a wide dynamic range, reflecting the broad range of BLR velocities covered by the sample, with the Gaussians scheme generally yielding larger values than the Lorentzian scheme and the KS comparison yielding a maximum separation of $D=0.1149$ between the two empirical cumulative distribution functions and a Monte Carlo $p$-value of $p_{\rm MC}<10^{-4}$.
This behavior is consistent with the expectation that a Lorentzian profile can describe a sharply peaked broad line with extended wings, which naturally leads to a smaller measured FWHM, whereas the Gaussians scheme tends to redistribute part of the broad-line flux into a broader component (see Figures~\ref{fig:fitexamples} and~\ref{fig:fitexample_ha}).
The broad-{\hb} luminosity distributions from the two fitting schemes are very similar, indicating that the broad-{\hb} luminosity is less sensitive to the adopted profile prescription than the FWHM. 
The {\feii} strength, expressed as
\begin{equation}
\mathcal{R}_{\rm Fe} = \frac{F_{\rm Fe~{\textsc{ii}},\,4434-4684\AA}}{F_{\mathrm{broad\,H\beta}}},
\end{equation}
also shows a modest but statistically detectable dependence on the broad-{\hb} decomposition ($D=0.0593$, $p_{\rm MC}<10^{-4}$).
This is expected because the {\feii} pseudo-continuum, the underlying AGN power-law continuum, and the broad-{\hb} wings are blended, so that changes in the adopted broad-{\hb} profile can redistribute flux among these components.

\subsection{Narrow Emission Lines}
\label{subsec:nel}

As shown in Figure~\ref{fig:dist-fitl2dg}, the {\oiii} luminosity distributions measured from the Lorentzian and Gaussians fitting schemes are highly consistent, as the {\oiii} doublet is modeled with the same narrow-line prescription in both fitting schemes and is relatively well separated from the broad~{\hb} component.
Therefore, the $L_{\rm [O\,\textsc{iii}]\lambda5007}$ measurement is insensitive to the adopted broad-{\hb} profile.
In contrast, the narrow~{\hb} luminosity shows a stronger profile dependence than {\oiii}, even though its fitting profile is tied to {\oiii}, because the two fitting schemes can assign different fractions of the total {\hb} flux to the broad and narrow components (as mentioned in Section~\ref{subsec:bel}).
This difference propagates into line ratios involving narrow~{\hb}, such as {\oiii}$\lambda5007$/narrow~{\hb}.

\begin{figure*}
\centering
\includegraphics[width=\textwidth]{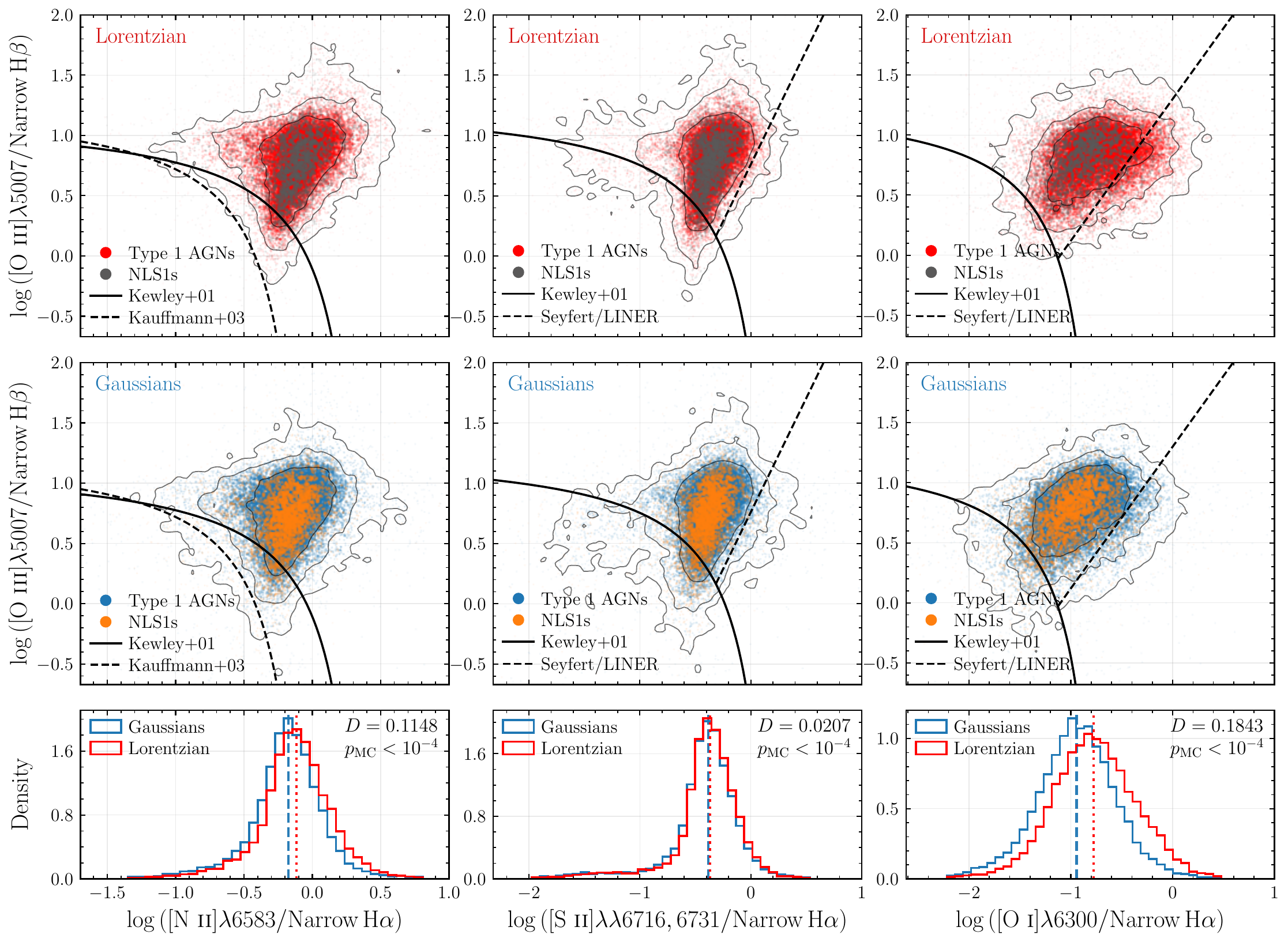}
\caption{
BPT diagnostic diagrams and line-ratio distributions for our type~1 AGN sample at $z\le0.35$.
The upper and middle rows show the BPT diagrams from the Lorentzian and Gaussians fitting schemes, respectively.
From left to right, the panels show the {\nii}, {\sii}, and {\oi} diagnostics.
Blue and red points show the full type~1 sample in the Gaussians and Lorentzian schemes, respectively, while orange and gray points mark the corresponding NLS1 subsets.
Contours show the density distribution of the full type~1 sample.
The solid curves show the theoretical maximum starburst boundaries from~\citet{Kewley2001}.
The dashed curve in the {\nii} diagram shows the empirical separation from~\citet{Kauffmann2003b}, while the dashed straight lines in the {\sii} and {\oi} diagrams indicate the Seyfert/LINER divisions from~\citet{Kewley2006}.
The bottom row compares the normalized distributions of the corresponding line ratios obtained from the two fitting schemes;
the median values, KS statistic $D$, and $p_{\rm MC}$ are also shown.
}
\label{fig:bpt}
\end{figure*}

Figure~\ref{fig:bpt} shows the {\nii}, {\sii}, and {\oi} BPT diagrams, together with the corresponding line-ratio distributions, for objects at $z\le0.35$ under the Lorentzian and Gaussians fitting schemes.
The global BPT distributions are broadly similar between the two fitting schemes, and most objects occupy the AGN/composite regions rather than the star-forming sequence.
Apart from the small vertical offsets caused by the profile dependence of narrow {\hb}, the most visible profile-dependent differences appear in the {\nii} and {\oi} BPT diagrams (with $D=0.1148$ and 0.1843 for the {\nii} and {\oi} line-ratio distributions, respectively, and $p_{\rm MC}<10^{-4}$ for both), especially for the {\oi}$\lambda6300$/narrow~{\ha} ratio, where the Lorentzian scheme gives larger values than the Gaussians scheme.
This offset is likely driven by differences in the narrow-{\ha} decomposition (see Figure~\ref{fig:fitexample_ha}), like in the {\hb} region.
Overall, the BPT classifications and narrow-line ratio distributions are broadly consistent between the two fitting schemes, but ratios involving narrow Balmer components can still show profile-dependent differences.

\subsection{Luminosity Relations}
\label{subsec:luminosity}

\begin{figure*}[!ht]
\centering
\includegraphics[width=\textwidth]{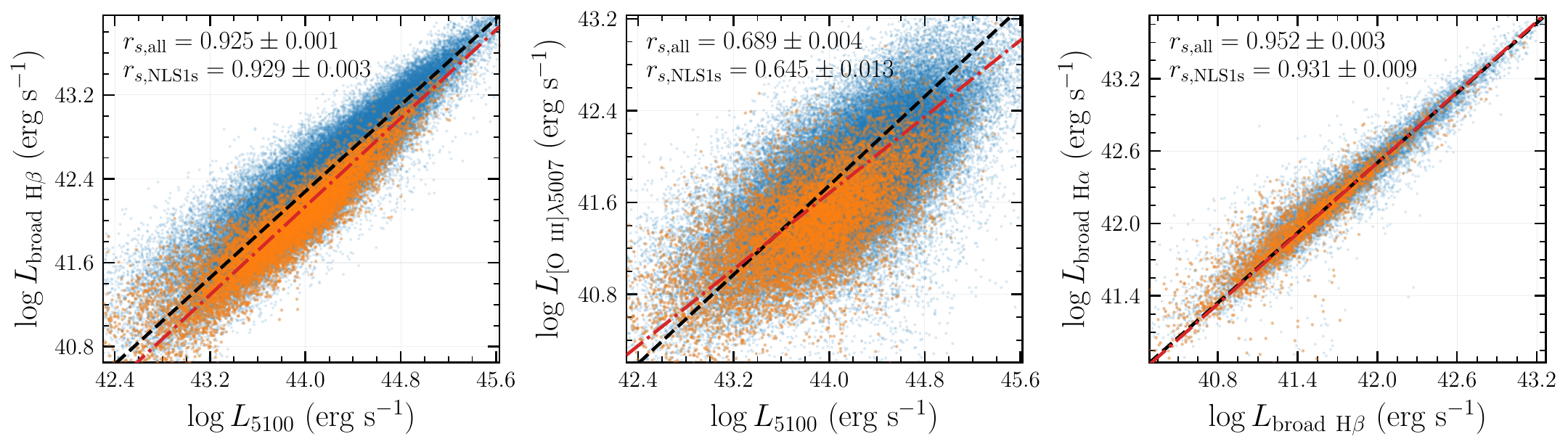}
\caption{
Luminosity relations for our type~1 AGN sample.
The left panel shows broad~{\hb} luminosity versus the host-subtracted 5100~{\AA} continuum luminosity.
The middle panel shows {\oiii}$\lambda5007$ luminosity versus $L_{5100}$.
The right panel compares broad~{\ha} and broad~{\hb} luminosities for the $z\le0.35$ subsample.
Blue points and black dashed lines represent the full type~1 sample and its best-fitting relation, respectively, while orange points and red dashed lines represent the NLS1 subset and its best-fitting relation, respectively.
}
\label{fig:rel-luminosities}
\end{figure*}

We first calculate the luminosity distributions of the main spectral components, including the host-subtracted continuum, broad Balmer emission, and narrow-line emission, as shown in Figure~\ref{fig:dist-fitl2dg}.
Then we examine the luminosity relations between the emission-line components and the continuum, as shown in Figure~\ref{fig:rel-luminosities}.
Specifically, Figure~\ref{fig:rel-luminosities} compares broad~{\hb} luminosity with $L_{5100}$, {\oiii}$\lambda5007$ luminosity with $L_{5100}$, and broad~{\ha} luminosity with broad~{\hb} luminosity for the lower-redshift {\ha} subsample.
The best-fitting relations and Spearman rank coefficients are summarized in Table~\ref{tab:luminosity}.
The broad Balmer-line luminosities show tight scaling relations with both the optical continuum and each other.
Broad~{\hb} follows $L_{5100}$ with nearly linear slopes for both the full type~1 sample and the NLS1 subset, while broad~{\ha} is tightly correlated with broad~{\hb} in the $z\le0.35$ subsample.
These trends indicate that the broad Balmer lines provide mutually consistent tracers of the low-ionization BLR emission.
By contrast, {\oiii}$\lambda5007$ shows a looser relation with $L_{5100}$, especially for the NLS1 subset.
This broader scatter and shallower NLS1 slope suggest that the NLR luminosity is not simply an instantaneous linear tracer of the observed optical continuum, but is also affected by the NLR covering factor, gas distribution, extinction, ionization structure, and the longer response timescale of the extended narrow-line gas.
This BLR--NLR contrast motivates the equivalent-width discussion in Section~\ref{subsec:ew}.

\begin{table*}
\centering
\caption{
Best-fitting relations among continuum and emission-line luminosities.
The fitted relation is written as $y=ax+b$, where $x$ and $y$ are the logarithmic luminosities listed in the first two columns.
}
\label{tab:luminosity}
\begin{tabular}{llcccl}
\hline
$y$ & $x$ & $a$ & $b$ & $r_{\rm s}$ & Sample \\
\hline
$\lg L_{\mathrm{H}\beta}$ & $\lg L_{5100}$ & $1.029\pm0.001$ & $-3.014\pm0.037$ & 0.925 & All type~1 \\
& & $1.053\pm0.002$ & $-4.178\pm0.107$ & 0.929 & NLS1 \\
\hline
$\lg L_{\mathrm{[O~{\textsc{iii}}]}\lambda5007}$ & $\lg L_{5100}$ & $0.970\pm0.002$ & $-0.930\pm0.071$ & 0.689 & All type~1 \\
& & $0.832\pm0.005$ & $5.091\pm0.208$ & 0.645 & NLS1 \\
\hline
$\lg L_{\mathrm{H}\alpha}$ & $\lg L_{\mathrm{H}\beta}$ & $0.966\pm0.001$ & $1.940\pm0.045$ & 0.952 & All type~1 \\
& & $0.975\pm0.002$ & $1.568\pm0.098$ & 0.931 & NLS1 \\
\hline
\end{tabular}
\end{table*}

\subsection{Black Hole Masses and Accretion Rates}
\label{subsec:mbh_mdot}

We estimate the black hole masses using the single-epoch virial method based on the broad~{\hb} line width and the host-subtracted 5100~{\AA} continuum luminosity. 
Assuming that the BLR gas is virialized, the black hole mass can be written as
\begin{equation}
    M_{\rm BH}=f\,\frac{R_{\rm BLR} \Delta V^2}{G},
    \label{eq:mbh_virial}
\end{equation}
where $R_{\rm BLR}$ is the BLR radius, $\Delta V$ is the velocity width of the BLR clouds, commonly represented by the FWHM of the broad~{\hb} line, and $f$ is the virial factor that depends on the geometry and kinematics of the BLR.
In this work, we adopt $f=3/4$, corresponding to the commonly used assumption of a spherical distribution of BLR clouds~\citep{Shen2013,Rakshit2017}.
We note that $f$ is not a fixed physical constant, but can depend on BLR geometry, inclination, and optical spectral properties; therefore, the black-hole masses derived here should be regarded as homogeneous single-epoch estimates rather than object-specific dynamical masses~\citep{HoKim2014,Pancoast2014,Yu2019,Yang2024}.
For the fiducial estimate, we use the $R_{\rm H\beta}$--$L_{5100}$ relation including the {\rfe} correction,
\begin{equation}
\log \left( \frac{R_{\rm H\beta}}{{\rm lt\mbox{-}day}} \right) = \alpha + \beta \log \ell_{44} + \gamma \mathcal{R}_{\rm Fe},
\label{eq:rblr_rfe}
\end{equation}
where $\ell_{44}=L_{5100}/(10^{44}~{\rm erg~s^{-1}})$.
The best-fitting coefficients are $\alpha=1.65\pm0.06$, $\beta=0.45\pm0.03$, and $\gamma=-0.35\pm0.08$~\citep{DuWang2019}.
The dimensionless mass accretion rate is estimated following the standard thin-disk prescription~\citep{ShakuraSunyaev1973,Du2015,Wang2014},
\begin{equation}
\dot{\mathscr{M}} = 20.1\left(\frac{\ell_{44}}{\cos i}\right)^{3/2}M_{7}^{-2},
\label{eq:mdot}
\end{equation}
where $M_7=M_{\rm BH}/10^7~M_{\odot}$.
The parameter $i$ is the inclination angle between the accretion disk and the line of sight.
We adopt $\cos i=0.75$, corresponding to a mean disk inclination for type~1 AGNs~\citep{Du2016b,Hu2020a,Li2021,Yao2024}.
The uncertainties in $R_{\rm BLR}$ are propagated from the measurement uncertainty of $L_{5100}$, the uncertainty of {\rfe}, and the quoted uncertainties of the coefficients in the radius--luminosity relation.
The uncertainties in {\mbh} are then propagated from the BLR radius and broad-{\hb} FWHM uncertainties, while those in {\mdot} are propagated from the uncertainties in $L_{5100}$ and {\mbh}.

Figure~\ref{fig:dist-fitl2dg} shows the distributions of $\log M_{\rm BH}$ and $\log \dot{\mathscr{M}}$ obtained from the Lorentzian and Gaussians schemes.
The two fitting schemes yield broadly similar distributions, indicating that the global range of virial masses and accretion rates is not strongly altered by the adopted broad-line profile.
However, the Gaussians scheme gives a black-hole mass distribution that is shifted slightly toward larger values compared with the Lorentzian scheme, following naturally from its larger broad-{\hb} FWHM measurements.
The dimensionless accretion-rate distribution shows the corresponding opposite behavior, with the Gaussians scheme placing sources at lower {\mdot} because a larger {\mbh} reduces the inferred accretion rate at fixed $L_{5100}$.
This profile-dependent shift is most relevant near the low-FWHM end, where modest differences in the broad-{\hb} decomposition can produce appreciable changes in {\mbh} and {\mdot}.

\section{Physical Implications of Profile-dependent Measurements}
\label{sec:physical_implications}

\subsection{The \texorpdfstring{$\mathcal{R}_{\rm Fe}$}{RFe}--FWHM Sequence}

\begin{figure*}
\centering
\includegraphics[width=0.96\textwidth]{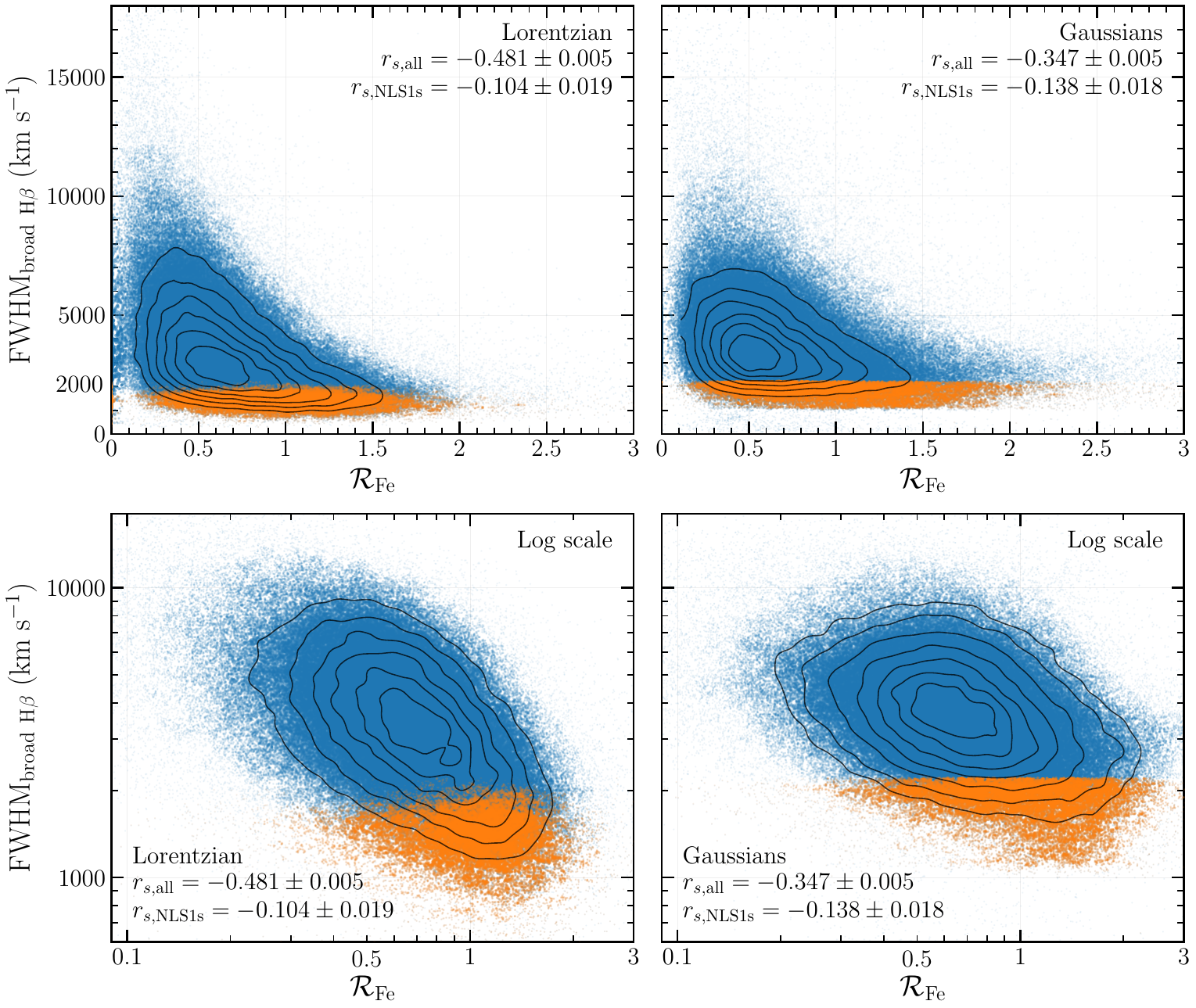}
\caption{
Relation between {\rfe} and the FWHM of broad {\hb} for the Lorentzian and Gaussians fitting schemes.
The left and right columns show the Lorentzian and Gaussians schemes, respectively.
The upper panels show the relation in linear coordinates, while the lower panels show the same relation in logarithmic coordinates.
Orange points denote the NLS1 subset, while blue points show the full type~1 sample, and black contours represent the density distribution of the full type~1 sample.
Spearman rank coefficients and their 95\% confidence intervals for the corresponding samples are shown in each panel.
}
\label{fig:rel-rfe-fwhmbhb}
\end{figure*}

The comparison between the Lorentzian and Gaussians fitting schemes allows us to examine which physical trends are robust against the adopted broad-{\hb} profile prescription. 
The {\rfe}--FWHM$_{\rm broad\,H\beta}$ plane is closely related to the quasar main-sequence, in which optical {\feii} strength, broad-{\hb} width, and {\oiii} strength are linked to the accretion state and other BLR properties\citep{BorosonGreen1992,Sulentic2000,SulenticMarziani2015,ShenHo2014,Marziani2018,Panda2019}.
Figure~\ref{fig:rel-rfe-fwhmbhb} shows the relation between {\feii} strength and broad-{\hb} FWHM for the two fitting schemes.
We present the relation in the conventional linear form and also in logarithmic coordinates, where the continuity of the {\rfe}--FWHM sequence becomes more clearly visible, especially when {\rfe} is shown on a logarithmic scale.

The full type~1 sample shows a clear anti-correlation between {\rfe} and broad-{\hb} FWHM in both fitting schemes.
The Spearman rank coefficient is $r_{\rm s}=-0.481$ for the Lorentzian scheme and $r_{\rm s}=-0.347$ for the Gaussians scheme.
Thus, the EV1-like trend, in which stronger optical {\feii} emission is associated with narrower broad {\hb}, is recovered regardless of the adopted broad-line profile prescription.
The correlation is stronger in the Lorentzian scheme, because the Lorentzian profile tends to yield smaller FWHM for sharply peaked broad~{\hb} profiles with extended wings.
By contrast, the Gaussians scheme shifts part of the low-FWHM population toward larger line widths, making the global anti-correlation less pronounced.
The larger broad-{\hb} FWHMs obtained with the Gaussians scheme are qualitatively consistent with the comparison presented by~\citet{Bian2008}.
For their NLS1 subset requiring two Gaussian components, they found that previously published Lorentzian-based FWHMs were $0.84\pm0.01$ times their two-Gaussian measurements, corresponding to virial masses lower by approximately 0.15~dex at fixed BLR radius and virial factor.
Moreover, in both fitting schemes, the logarithmic {\rfe}–FWHM diagram shows a continuous distribution from low-FWHM, strong-{\feii} objects to broader-line, weaker-{\feii} type~1 AGNs, which indicates that NLS1s occupy one end of the type~1 sequence in this plane rather than forming a distinct population.

\subsection{NLS1 Classification}
\label{subsec:nls1_class}

Classical NLS1 definitions are based on empirical optical criteria, most importantly the small width of the Balmer lines, together with strong optical {\feii} emission and relatively weak narrow-line emission~\citep{OsterbrockPogge1985,Goodrich1989,Komossa2008b}. 
Large SDSS NLS1 catalogs~\citep{Zhou2006,Rakshit2017,Paliya2024} have commonly adopted a broad-{\hb} FWHM threshold of about $2000$~{\kms}, although the exact classification can depend on spectral decomposition and data quality.
Thus, the NLS1 classification is sensitive to the adopted broad-line profile prescription.
In this work, we adopt criteria that include the FWHM uncertainty:
\begin{equation}
{\rm FWHM}_{\rm broad\,H\beta} - {\rm FWHM\ Err}_{\rm broad\,H\beta} < 2000~{\rm km~s^{-1}},
\end{equation}
and
\begin{equation}
{\rm FWHM}_{\rm broad\,H\beta} < 2200~{\rm km~s^{-1}}.
\end{equation}
Using these criteria, we identify 25,628 NLS1s from the Lorentzian fitting scheme and 13,359 NLS1s from the Gaussians fitting scheme.
The intersection of the two selections contains 11,012 objects, which we define as the NLS1 subset used for the type~1--NLS1 comparison analysis.
Therefore, the comparison between the two fitting schemes shows that the classical NLS1/broad-line Seyfert~1 (BLS1) classification is profile-dependent, with objects close to the $2000$~{\kms} FWHM boundary classified as NLS1s in one fitting scheme but as BLS1s in the other.
This demonstrates that the empirical NLS1/BLS1 boundary should not be interpreted as a sharp physical division.

\subsection{The Accretion Sequence}
\label{subsec:accretion_sequence}

\begin{figure*}
\centering
\includegraphics[width=0.96\textwidth]{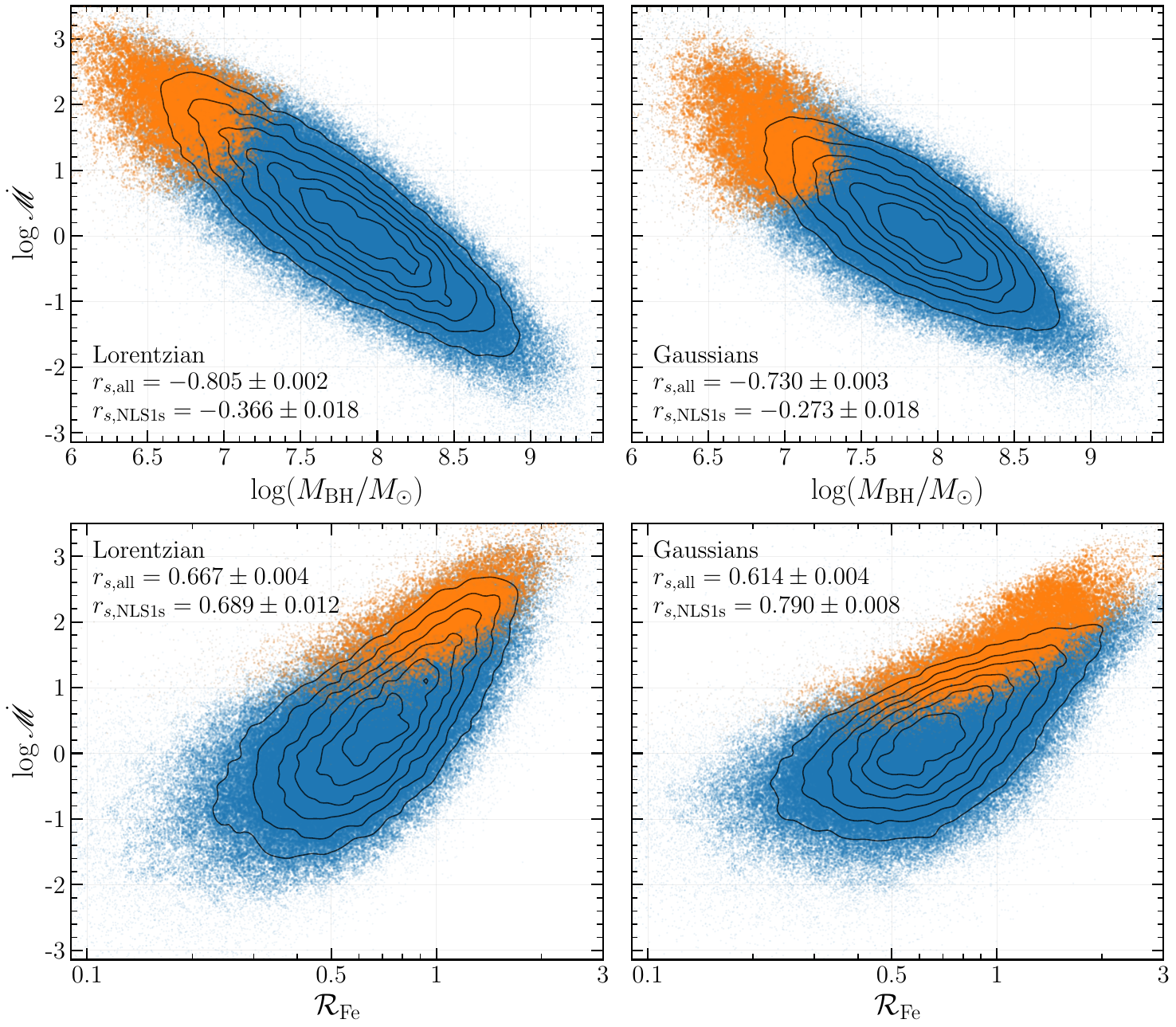}
\caption{
Relations among virial black-hole mass, dimensionless accretion rate, and {\rfe}.
The left and right columns show the results from the Lorentzian and Gaussians fitting schemes, respectively.
The top row shows the {\mbh}–{\mdot} relation, and the bottom row shows the {\rfe}–{\mdot} relation.
Orange points denote the NLS1 subset, while blue points show the full type~1 sample.
Black contours represent the density distribution of the full type~1 sample.
Spearman rank coefficients for the corresponding samples are shown in each panel.
}
\label{fig:rel-mbhmdotrfe}
\end{figure*}

Figure~\ref{fig:rel-mbhmdotrfe} shows the {\mbh}--{\mdot} and {\rfe}--{\mdot} relations for the Lorentzian and Gaussians fitting schemes.
For the full type~1 sample, {\mbh} and {\mdot} show a strong anti-correlation in both fitting schemes, with $r_{\rm s}=-0.805$ for the Lorentzian scheme and $r_{\rm s}=-0.730$ for the Gaussians scheme.
This strong anti-correlation is expected from the definition of the dimensionless accretion rate, {\mdot}~$\propto$~{\mbh}$^{-2}$, which directly links {\mdot} to {\mbh} at fixed optical luminosity.
The NLS1 subset follows the same overall trend but shows a weaker internal anti-correlation, with $r_{\rm s}=-0.366$ and $-0.273$ for the Lorentzian and Gaussians schemes, respectively.
This weaker correlation is likely related to the fact that NLS1s are concentrated toward the low-{\mbh} and high-{\mdot} end of the full type~1 distribution, rather than being uniformly distributed along the entire {\mbh}--{\mdot} sequence.
This location is consistent with the commonly discussed picture that NLS1s tend to have smaller virial black-hole masses and higher inferred accretion rates~\citep{Pounds1995,Mathur2000,CollinKawaguchi2004,Komossa2008b,Wang2013,Du2016a}.

The {\rfe}--{\mdot} relation shows a positive correlation in both fitting schemes.
For the full type~1 sample, the Spearman coefficient is $r_{\rm s}=0.667$ for the Lorentzian scheme and $r_{\rm s}=0.614$ for the Gaussians scheme.
This result is consistent with the commonly discussed connection between strong optical {\feii} emission and high inferred accretion rate.
However, {\rfe} is not a single-parameter accretion-rate tracer, because it can also depend on the ionizing SED, gas density, column density, metallicity, microturbulence, and orientation~\citep{Panda2019,Panda2024}.
In addition, the {\rfe}--{\mdot} relation is not fully independent of the adopted virial prescription, because {\rfe} enters the adopted $R_{\rm H\beta}$--$L_{5100}$ relation used to estimate {\mbh} and {\mdot}.
These additional dependencies may partly explain why the {\rfe}--{\mdot} correlation is weaker than the built-in {\mbh}--{\mdot} anti-correlation for the full type~1 sample.
For the NLS1 subset, however, the {\rfe}--{\mdot} correlation is stronger, with $r_{\rm s}=0.689$ and $0.790$ for the Lorentzian and Gaussians schemes, respectively.
This may be because, within the low-FWHM occupied by NLS1s, variations in {\rfe} more directly trace differences in the inferred accretion state, while the restricted FWHM range reduces the dynamic range driven by line width alone.

Overall, the two broad-{\hb} fitting schemes change the numerical values of {\mbh} and {\mdot}, but they do not change the qualitative accretion sequence.
Stronger optical {\feii} emission is associated with higher inferred accretion rates, and NLS1s are located toward the high-accretion end of the type~1 population.

\subsection{Equivalent-width Correlations}
\label{subsec:ew}

\begin{figure*}
\centering
\includegraphics[width=0.96\textwidth]{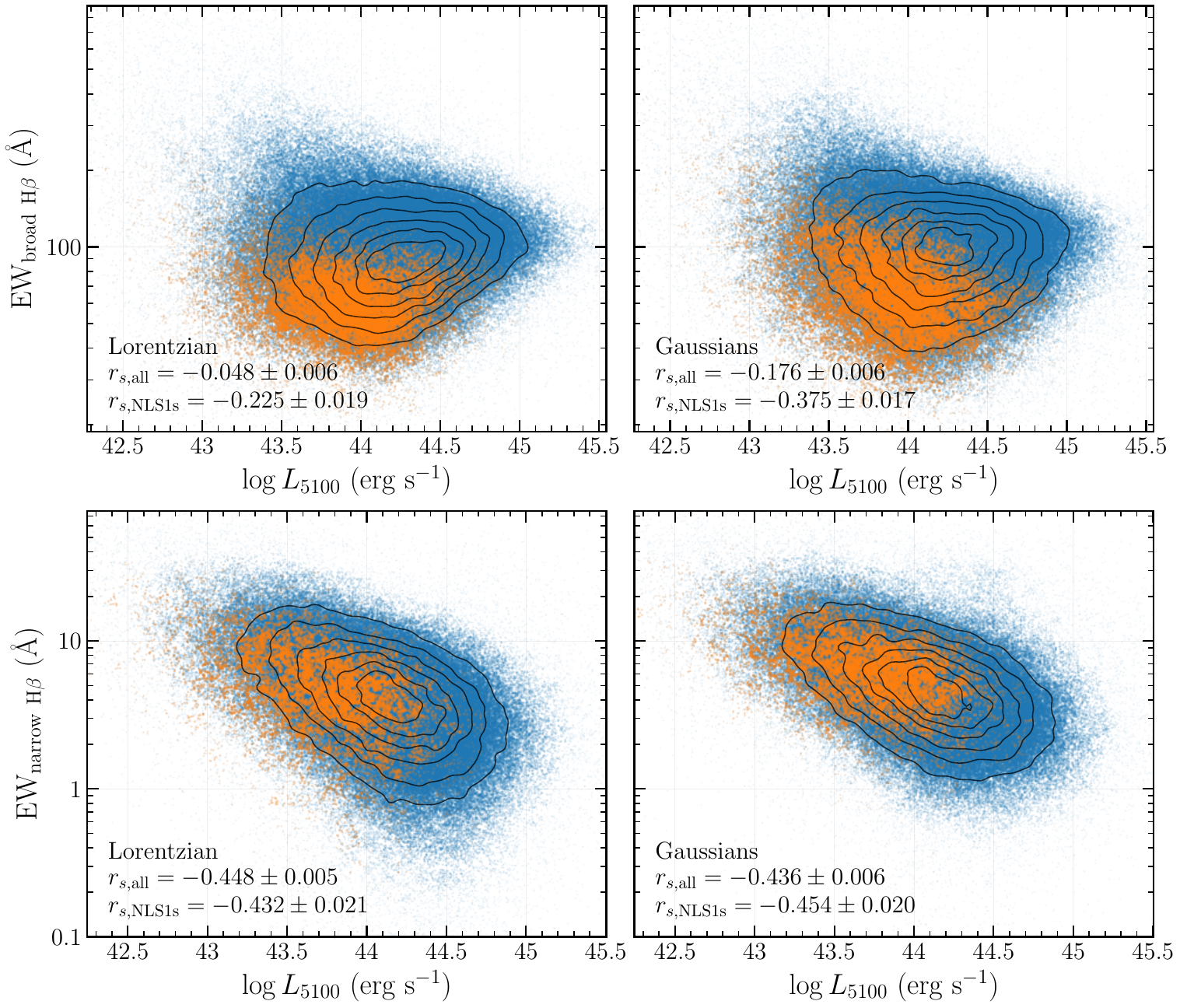}
\caption{
Relations between the {\hb} equivalent widths and the host-subtracted 5100~{\AA} continuum luminosity.
The left and right columns show the results from the Lorentzian and Gaussians fitting schemes, respectively.
The upper and lower rows show EW$_{\rm broad\,H\beta}$ and EW$_{\rm narrow\,H\beta}$, respectively.
Orange points denote NLS1 subset, while blue points represent the full type~1 AGNs.
Black contours show the density distribution of the full type~1 sample.
Spearman rank coefficients for the corresponding samples are shown in each panel.
}
\label{fig:ew_l5100}
\end{figure*}

We examine the relations between the {\hb} equivalent widths (EW) and the host-subtracted continuum luminosity $L_{5100}$.
Equivalent-width trends with continuum luminosity are commonly discussed in the context of the Baldwin effect, originally identified for broad UV lines and later extended to optical and narrow emission lines~\citep{Baldwin1977,Baldwin1989,Dietrich2002,Croom2004,Zhang2013}. 
For narrow lines, previous studies have shown that their relative strength can decrease with AGN luminosity, possibly reflecting a luminosity-dependent NLR covering factor, continuum anisotropy, variability, or changes in the ionizing SED~\citep{BaskinLaor2005,SternLaor2012a,Zhang2013}.
Figure~\ref{fig:ew_l5100} shows the equivalent widths of broad and narrow {\hb} as functions of $L_{5100}$ for the Lorentzian and Gaussians fitting schemes.

The equivalent width of broad~{\hb} shows only a weak anti-correlation with $L_{5100}$ in the full type~1 sample.
The Spearman rank coefficients are $r_{\rm s}=-0.048$ and $-0.176$ for the Lorentzian and Gaussians schemes, respectively.
The anti-correlation is more pronounced in the NLS1 subset, with $r_{\rm s}=-0.225$ for the Lorentzian scheme and $r_{\rm s}=-0.375$ for the Gaussians scheme.
Thus, although the global EW$_{\rm broad\,H\beta}$--$L_{5100}$ relation is weak, its strength depends to some extent on the adopted broad-{\hb} fitting scheme, particularly for NLS1s.
In comparison, EW$_{\rm narrow\,H\beta}$ exhibits a clearer anti-correlation with $L_{5100}$.
For the full type~1 sample, the Spearman coefficients are $r_{\rm s}=-0.448$ and $-0.436$ for the Lorentzian and Gaussians schemes, respectively.
Similar correlations are found for the NLS1 subset, with $r_{\rm s}=-0.432$ and $-0.454$, respectively.
The close agreement between the two fitting schemes indicates that the EW$_{\rm narrow\,H\beta}$--$L_{5100}$ anti-correlation is qualitatively insensitive to the adopted broad-{\hb} profile, although the individual narrow-{\hb} measurements can still be affected by the decomposition of the {\hb} complex.

Overall, the equivalent-width analysis provides a complementary view of the profile-dependent measurements.
Both fitting schemes recover a weak EW$_{\rm broad,H\beta}$--$L{5100}$ relation but a much stronger EW$_{\rm narrow,H\beta}$--$L{5100}$ anti-correlation, indicating that the broad and narrow Balmer components respond differently to continuum luminosity.
This contrast is present in both the Lorentzian and Gaussians schemes, so it is not driven by the adopted broad-{\hb} profile prescription.
As for the NLS1 subset, it shows a stronger EW$_{\rm broad\,H\beta}$--$L_{5100}$ anti-correlation than the full type~1 sample, but in all panels the NLS1s remain located toward one corner of the overall type~1 distribution in these relations rather than forming a distinct population.

\subsection{NLR Excitation along the Accretion-state Sequence}
\label{subsec:nlr_excitation}

\begin{figure*}[!ht]
\centering
\includegraphics[width=0.96\textwidth]{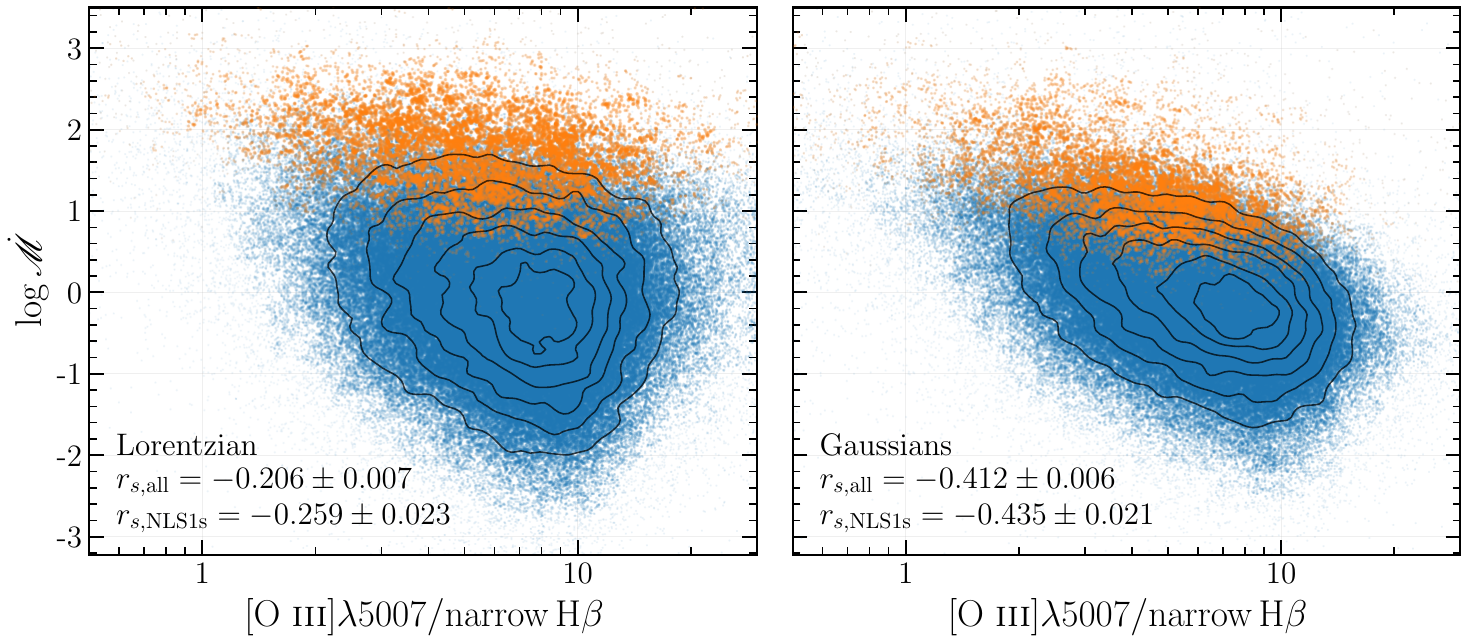}
\caption{
Relation between the {\oiii}$\lambda5007$/narrow~{\hb} and the dimensionless accretion rate. 
The left and right panels show the results from the Lorentzian and Gaussians fitting schemes, respectively.
Orange points denote the NLS1 subset, while blue points represent the full type~1 AGNs.
Black contours show the density distribution of the full type~1 sample.
Spearman rank coefficients for the corresponding samples are shown in each panel.
}
\label{fig:rel-mdotoiiinhb}
\end{figure*}

The narrow-line ratio {\oiii}$\lambda5007$/narrow~{\hb} is commonly used as an excitation indicator for the NLR gas and is sensitive to the ionization parameter, gas density, metallicity, extinction, and the shape and anisotropy of the ionizing continuum~\citep{Veilleux1987,Kewley2006,BaskinLaor2005,SternLaor2012b}.
It therefore provides a complementary view of the type~1 sequence on larger spatial scales.
Figure~\ref{fig:rel-mdotoiiinhb} shows the relation between {\oiii}$\lambda5007$/narrow~{\hb} and $\log \dot{\mathscr M}$ for the Lorentzian and Gaussians fitting schemes.
The full sample shows an anti-correlation in both fitting schemes, with $r_{\rm s}=-0.206$ for the Lorentzian scheme and $r_{\rm s}=-0.412$ for the Gaussians scheme. 
The NLS1 subset shows the same qualitative trend, with $r_{\rm s}=-0.259$ and $-0.435$, respectively.

This anti-correlation indicates that objects with higher inferred accretion rates tend to have lower {\oiii}$\lambda5007$/narrow~{\hb} ratios. 
It suggests that the high-accretion end of the type~1 population is associated with relatively weaker {\oiii} emission with respect to narrow~{\hb}.
This behavior is consistent with the EV1 picture in which strong {\feii}, narrow broad {\hb}, high inferred accretion rate, and weaker {\oiii} emission tend to occur together.

Several physical effects may contribute to this trend. 
A softer or more strongly filtered ionizing continuum in high-accretion systems can reduce the production efficiency of high-ionization narrow lines, while disk winds or geometrically thick inner accretion flows may shield part of the NLR from the hardest ionizing photons~\citep{LeighlyMoore2004,Luo2015,Panda2019}. 
In addition, the observed {\oiii} strength depends on the effective NLR covering factor, gas distribution, extinction, and orientation-dependent continuum anisotropy~\citep{BaskinLaor2005,Risaliti2011,SternLaor2012b,Zhang2013}. 
Narrow~{\hb} can also be affected by host-galaxy or star-forming contamination, especially in lower-luminosity or host-dominated spectra.
Thus, the {\oiii}/narrow~{\hb}--{\mdot} relation should be interpreted as an empirical coupling between NLR excitation and the BLR accretion sequence, rather than as a direct causal relation between the narrow-line ratio and instantaneous mass accretion rate.

The profile dependence is strongest for quantities that depend directly on broad-{\hb} FWHM, including {\mbh}, {\mdot}, and the low-FWHM classification. 
Therefore, broad-{\hb} profile assumptions do not erase the physical organization of the type~1 population, but they do affect the quantitative mapping from observed line profiles to virial physical parameters, which is broadly consistent with the recent host-galaxy-based results of NLS1~\citep{Guo2026}.

\section{Conclusion}
\label{sec:sum}

We construct a sample of 114,754 low-redshift type~1 AGNs from DESI~DR1 through host-galaxy decomposition, emission-line fitting, and ML-assisted sample refinement.
Each spectrum is analyzed using both Lorentzian and Gaussians broad-Balmer fitting schemes, while all other spectral components and fitting constraints are treated identically.
Our main conclusions are as follows.

\begin{enumerate}

\item
The spectral measurements show clear dependence on the adopted broad-Balmer profile.
The distributions show that the Gaussians scheme generally yields larger broad-{\hb} FWHM values than the Lorentzian scheme, especially at the low-FWHM end, and consequently gives larger virial black-hole masses and lower dimensionless accretion rates.
The Gaussians scheme also gives larger {\rfe}, while narrow-line ratios involving narrow {\hb} or narrow {\ha} are generally smaller than those from the Lorentzian scheme because of the different broad--narrow Balmer decomposition.
By contrast, the measured broad-{\hb} luminosity and {\oiii}$\lambda5007$ luminosity remain comparatively stable between the two schemes.

\item
The adopted broad-Balmer profile also affects the statistical correlations in the {\rfe}--FWHM plane, the {\mbh}--{\mdot} plane, the {\rfe}--{\mdot} plane, the EW--$L_{5100}$ relations, the BPT diagrams, and the {\oiii}~$\lambda5007$/narrow~{\hb}--{\mdot} relation, with the correlation strengths changing to different degrees.
The profile dependence is especially visible in quantities involving narrow Balmer emission, including the {\oi} and {\nii} BPT diagrams and the {\oiii}$\lambda5007$/narrow {\hb}--{\mdot} relation.
For the latter relation, the anti-correlation is substantially stronger in the Gaussians scheme than in the Lorentzian scheme, with Spearman rank correlation coefficients of $r_{\rm s}=-0.409$ and $-0.207$ for the full type~1 sample, and $r_{\rm s}=-0.352$ and $-0.258$ for the NLS1 subset, respectively.
These results favor the Gaussians scheme as a more self-consistent decomposition of the broad and narrow Balmer components.

\item
Despite these quantitative differences, the main relations are recovered under both fitting schemes.
Both fitting schemes reveal a continuous type~1 accretion sequence, whose continuity becomes more apparent in the logarithmic {\rfe}--FWHM plane and along which decreasing broad-{\hb} FWHM is accompanied by stronger optical {\feii} emission and higher accretion rates.
Along the same sequence, the {\oiii}$\lambda5007$/narrow~{\hb} ratio tends to decrease toward the high-{\mdot} end, indicating that the BLR accretion sequence is accompanied by systematic changes in narrow-line ratios.
For the objects classified as NLS1s under both fitting schemes, we find that they do not form a physically distinct population in these diagrams, but instead occupy one end of the continuous type~1 accretion sequence.

\end{enumerate}
Overall, broad-Balmer profile assumptions affect the quantitative measurements of BLR, narrow-line, and virial quantities, as well as the strengths of several statistical correlations.
However, they do not erase the overall organization of the type~1 population.
Therefore, broad-line profile prescriptions should be explicitly considered when interpreting single-epoch black-hole masses, accretion rates, narrow-line ratios, and NLS1 classifications.

\begin{acknowledgments}

We acknowledge the financial support from the National Natural Science Foundation of China under grants No. 12273054 and 12503019.

This material is based upon work supported by the U.S. Department of Energy (DOE), Office of Science, Office of High-Energy Physics, under Contract No. DE–AC02–05CH11231, and by the National Energy Research Scientific Computing Center, a DOE Office of Science User Facility under the same contract. Additional support for DESI was provided by the U.S. National Science Foundation (NSF), Division of Astronomical Sciences under Contract No. AST-0950945 to the NSF’s National Optical-Infrared Astronomy Research Laboratory; the Science and Technologies Facilities Council of the United Kingdom; the Gordon and Betty Moore Foundation; the Heising-Simons Foundation; the French Alternative Energies and Atomic Energy Commission (CEA); the National Council of Science and Technology of Mexico (CONACYT); the Ministry of Science and Innovation of Spain (MICINN), and by the DESI Member Institutions: \url{https://www.desi.lbl.gov/collaborating-institutions}.

Any opinions, findings, and conclusions or recommendations expressed in this material are those of the author(s) and do not necessarily reflect the views of the U. S. National Science Foundation, the U. S. Department of Energy, or any of the listed funding agencies.

The authors are honored to be permitted to conduct scientific research on Iolkam Du’ag (Kitt Peak), a mountain with particular significance to the Tohono O’odham Nation.
\end{acknowledgments}


\facilities{DESI}

\software{
\texttt{Astropy}~\citep{Astropy2013,Astropy2018,Astropy2022}, 
\texttt{DASpec}~\citep{Du2024_daspec},
\texttt{PyTorch}~\citep{Paszke2019},
\texttt{scikit-learn}~\citep{Pedregosa2011}
}

\appendix

\section{Catalog Description}
\label{app:catalog_columns}

The machine-readable catalog released with this work contains the final sample of 114,754 low-redshift broad-{\hb} type~1 AGNs selected from DESI~DR1.
For each object, we provide the basic DESI targetid, coordinates, redshift, sample-selection flags, ML-assisted refinement quantities, and two parallel sets of spectral measurements obtained from the Lorentzian and Gaussians fitting schemes.
The catalog also includes derived virial quantities and, for the $z\le0.35$ subsample, {\ha}-region measurements and BPT diagnostic ratios.
Both Lorentzian-based and Gaussians-based measurements are provided so that users can assess the systematic effects associated with the adopted broad-Balmer profile.
A summary of the main catalog columns is given in Table~\ref{tab:catalog_columns}; the full catalog will be made available in machine-readable format upon publication.

\begin{deluxetable*}{lll}
\tabletypesize{\scriptsize}
\tablewidth{0pt}
\tablecaption{Column description of the machine-readable catalog.
\label{tab:catalog_columns}}
\tablehead{
\colhead{Column} &
\colhead{Unit} &
\colhead{Description}
}
\startdata
\texttt{TARGETID} & -- & DESI target identifier. \\
\texttt{NLS1\_FLAG\_<prof>} & -- & Profile-specific NLS1 flag from the adopted broad-{\hb} FWHM criterion. \\
\texttt{NLS1\_FLAG\_BOTH} & -- & Flag for objects classified as NLS1 by both \texttt{fitl} and \texttt{fitg}. \\
\texttt{Z} & -- & DESI redshift used in this work. \\
\texttt{snr5100} & -- & Median continuum signal-to-noise ratio near rest-frame 5100~{\AA}. \\
\texttt{D4000} & -- & $D4000$ index measured from the available spectrum or product. \\
\texttt{Dn4000} & -- & Narrow $D_n(4000)$ index measured from the available spectrum or product. \\
\texttt{L5100} & erg s$^{-1}$ & Host-subtracted monochromatic luminosity, $\lambda L_{\lambda}$, at 5100~{\AA}. \\
\texttt{L5100\_err} & erg s$^{-1}$ & Uncertainty on \texttt{L5100}. \\
\texttt{relerr\_L5100} & -- & Relative uncertainty on \texttt{L5100}. \\
\texttt{BROAD\_PROFILE\_<prof>} & -- & Broad-{\hb} profile model used in the {\hb}-region fit. \\
\texttt{CHI2\_RED\_<prof>} & -- & Reduced chi-square of the {\hb}-region fit. \\
\texttt{Flux\_5100A\_<prof>} & fit flux unit & Power-law continuum flux density at rest-frame 5100~{\AA}. \\
\texttt{Flux\_5100A\_Err\_<prof>} & fit flux unit & Uncertainty on \texttt{Flux\_5100A\_<prof>}. \\
\texttt{PL\_Index\_5100A\_<prof>} & -- & Power-law continuum index. \\
\texttt{PL\_Index\_5100A\_Err\_<prof>} & -- & Uncertainty on \texttt{PL\_Index\_5100A\_<prof>}. \\
\texttt{Flux\_<component>\_<prof>} & fit flux unit & {\hb}-region component flux; \texttt{<component>} includes \texttt{FeII}, \texttt{Broad\_Hbeta}, \texttt{Narrow\_Hbeta}, \texttt{OIII4959}, and \texttt{OIII5007}. \\
\texttt{Flux\_<component>\_Err\_<prof>} & fit flux unit & Uncertainty on \texttt{Flux\_<component>\_<prof>}. \\
\texttt{FWHM\_<component>\_<prof>} & km s$^{-1}$ & {\hb}-region component FWHM. \\
\texttt{FWHM\_<component>\_Err\_<prof>} & km s$^{-1}$ & Uncertainty on \texttt{FWHM\_<component>\_<prof>}. \\
\texttt{Flux\_Broad\_Hbeta\_G1\_fitg} & fit flux unit & Flux of the first broad-{\hb} Gaussian component in the Gaussians fit. \\
\texttt{Flux\_Broad\_Hbeta\_G2\_fitg} & fit flux unit & Flux of the second broad-{\hb} Gaussian component in the Gaussians fit. \\
\texttt{FWHM\_Broad\_Hbeta\_G1\_fitg} & km s$^{-1}$ & FWHM of the first broad-{\hb} Gaussian component. \\
\texttt{FWHM\_Broad\_Hbeta\_G2\_fitg} & km s$^{-1}$ & FWHM of the second broad-{\hb} Gaussian component. \\
\texttt{Broad\_Hbeta\_G1\_Fraction\_fitg} & -- & Fractional contribution of the first broad-{\hb} Gaussian component. \\
\texttt{L\_OIII5007\_<prof>} & erg s$^{-1}$ & [O~{\sc iii}] $\lambda5007$ luminosity. \\
\texttt{EW\_Broad\_Hbeta\_<prof>} & {\AA} & Broad-{\hb} equivalent width. \\
\texttt{EW\_OIII5007\_<prof>} & {\AA} & [O~{\sc iii}] $\lambda5007$ equivalent width. \\
\texttt{EW\_OIII5007\_Err\_<prof>} & {\AA} & Uncertainty on \texttt{EW\_OIII5007\_<prof>}. \\
\texttt{RFe\_<prof>} & -- & Fe~{\sc ii} to broad-{\hb} flux ratio. \\
\texttt{RFe\_Err\_<prof>} & -- & Uncertainty on \texttt{RFe\_<prof>}. \\
\texttt{OIII5007\_to\_Narrow\_Hbeta\_<prof>} & -- & [O~{\sc iii}] $\lambda5007$ to narrow-{\hb} flux ratio. \\
\texttt{MBH\_<prof>} & $M_{\odot}$ & Single-epoch black-hole mass from broad-{\hb} FWHM and the adopted radius--luminosity prescription. \\
\texttt{mdot\_<prof>} & -- & Dimensionless accretion-rate proxy. \\
\texttt{CHI2\_RED\_Halpha\_<prof>} & -- & Reduced chi-square of the H$\alpha$-region fit. \\
\texttt{Flux\_5100A\_Halpha\_<prof>} & fit flux unit & Power-law continuum flux density at rest-frame 5100~{\AA} in the H$\alpha$-region fit. \\
\texttt{Flux\_<ha\_line>\_Halpha\_<prof>} & fit flux unit & H$\alpha$-region line flux; \texttt{<ha\_line>} includes \texttt{Broad\_Halpha}, \texttt{Narrow\_Halpha}, \texttt{NII6548}, \texttt{NII6583}, \texttt{SII6716}, \texttt{SII6731}, \texttt{OI6300}, and \texttt{OI6364}. \\
\texttt{Flux\_<ha\_line>\_Err\_Halpha\_<prof>} & fit flux unit & Uncertainty on \texttt{Flux\_<ha\_line>\_Halpha\_<prof>}. \\
\texttt{FWHM\_<ha\_line>\_Halpha\_<prof>} & km s$^{-1}$ & H$\alpha$-region line FWHM. \\
\texttt{FWHM\_<ha\_line>\_Err\_Halpha\_<prof>} & km s$^{-1}$ & Uncertainty on \texttt{FWHM\_<ha\_line>\_Halpha\_<prof>}. \\
\texttt{Flux\_Broad\_Halpha\_G1\_fitg} & fit flux unit & Flux of the first broad-H$\alpha$ Gaussian component in the Gaussians fit. \\
\texttt{Flux\_Broad\_Halpha\_G2\_fitg} & fit flux unit & Flux of the second broad-H$\alpha$ Gaussian component in the Gaussians fit. \\
\texttt{FWHM\_Broad\_Halpha\_G1\_fitg} & km s$^{-1}$ & FWHM of the first broad-H$\alpha$ Gaussian component. \\
\texttt{FWHM\_Broad\_Halpha\_G2\_fitg} & km s$^{-1}$ & FWHM of the second broad-H$\alpha$ Gaussian component. \\
\texttt{Broad\_Halpha\_G1\_Fraction\_fitg} & -- & Fractional contribution of the first broad-H$\alpha$ Gaussian component. \\
\texttt{<ratio>\_BPT\_<prof>} & -- & Linear BPT flux ratio from the matched {\hb}/H$\alpha$ measurements. \\
\enddata
\tablecomments{
The placeholder \texttt{<prof>} is expanded as \texttt{fitl} for Lorentzian fits and \texttt{fitg} for Gaussians fits.
Fluxes from the fitting tables are stored in fitted flux units; 
physical fluxes can be obtained by multiplying by the corresponding unit factor and by $10^{-17}${\ergscma}.
}
\end{deluxetable*}

\section{Additional Comparison between Lorentzian and Gaussians Measurements}
\label{app:compldg}

\begin{figure*}
\centering
\includegraphics[width=\textwidth]{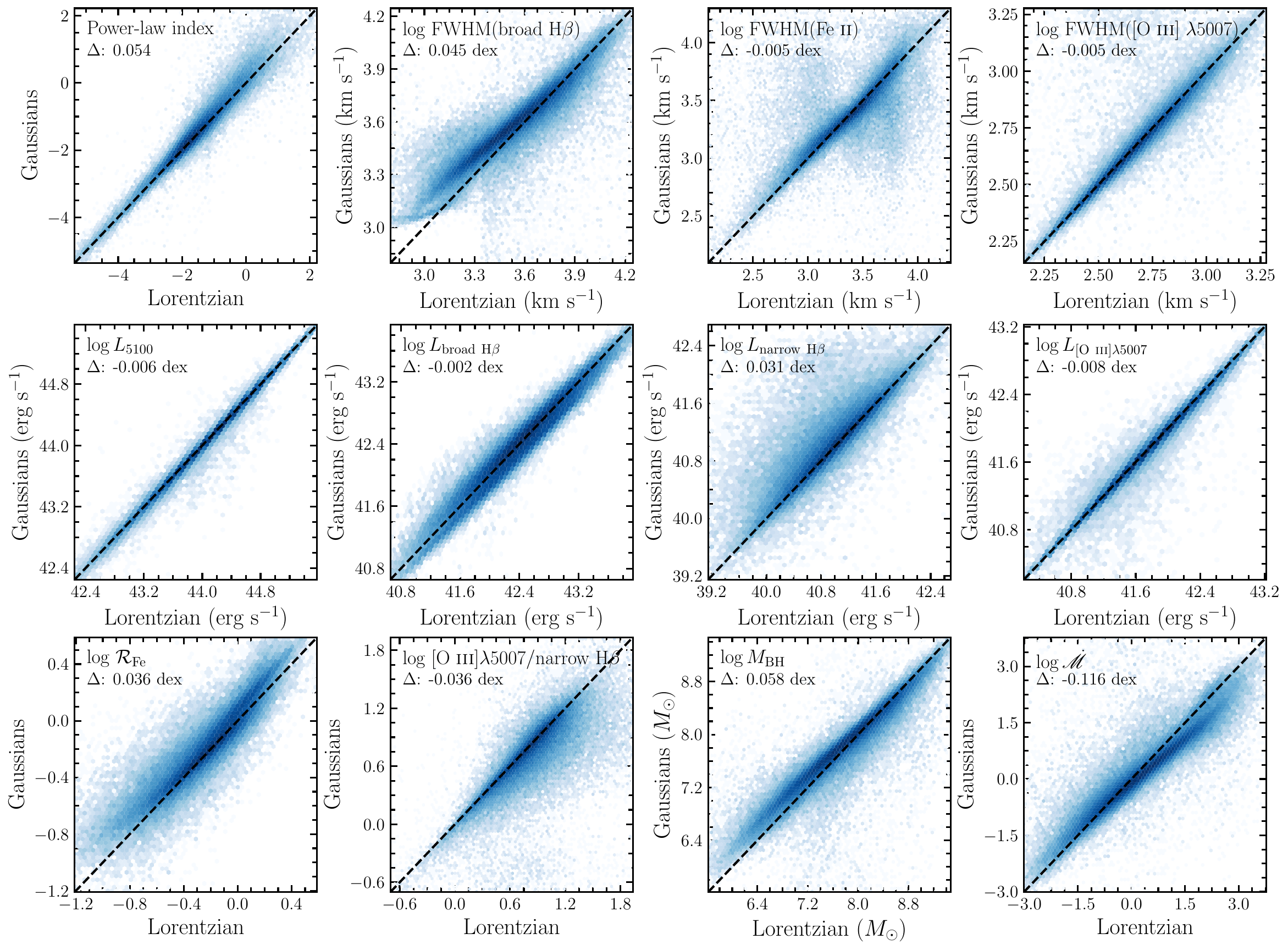}
\caption{
Direct comparison of the main quantities measured from the Lorentzian and Gaussians broad-{\hb} fitting schemes.
The horizontal axes show the Lorentzian-based measurements, and the vertical axes show the Gaussians-based measurements.
The panels show, from left to right and from top to bottom, the local AGN power-law index, broad-{\hb} FWHM, {\feii} FWHM, {\oiii}$\lambda5007$ FWHM, host-subtracted 5100~{\AA} luminosity, broad-{\hb} luminosity, narrow {\hb} luminosity, {\oiii}$\lambda5007$ luminosity, {\rfe}, {\oiii}$\lambda5007$/narrow~{\hb}, virial black-hole mass, and dimensionless accretion rate.
The black dashed line marks the one-to-one relation.
The median offset $\Delta$ is defined as the Gaussians-based value minus the Lorentzian-based value and is labeled in each panel.
}
\label{fig:appendix_compldg}
\end{figure*}

In this appendix, we provide a direct comparison of the main quantities measured from the Lorentzian and Gaussians fitting schemes.
This comparison is intended to show which measurements are stable against the adopted broad-{\hb} profile prescription and which measurements are more strongly affected by the decomposition of the {\hb} complex.
In Figure~\ref{fig:appendix_compldg}, the horizontal axes show the Lorentzian-based measurements, while the vertical axes show the Gaussians-based measurements.
The dashed line in each panel marks the one-to-one relation.

The continuum and narrow-line quantities show the smallest offsets between the two fitting schemes.
The host-subtracted continuum luminosity at 5100~{\AA} has a median offset of only $-0.006$~dex, and the {\oiii}$\lambda5007$ luminosity has a median offset of $-0.008$~dex.
The {\oiii}$\lambda5007$ FWHM is also nearly unchanged, with a median offset of $-0.005$~dex.
These small differences indicate that the local continuum and {\oiii} measurements are largely insensitive to the adopted broad-{\hb} profile prescription.
This is expected because these components are not primarily determined by the detailed shape of the broad {\hb} line.

The broad-{\hb} FWHM shows a larger systematic difference.
The Gaussians-based measurements are shifted to larger broad-{\hb} widths, with a median offset of $0.045$~dex in $\log {\rm FWHM}_{\rm broad\,H\beta}$.
This offset reflects the different ways in which the two empirical profiles describe sharply peaked broad lines with extended wings.
A Lorentzian profile can describe such profiles with a smaller measured FWHM, whereas the Gaussians scheme tends to assign part of the broad-line flux to a broader Gaussian component.
This difference is the main source of the profile dependence discussed in the main text.

The {\feii} and narrow {\hb} measurements show intermediate behavior.
The {\feii} FWHM is nearly stable, with a median offset of $-0.005$~dex, but {\rfe} is shifted by $0.036$~dex.
This indicates that the {\feii} template broadening itself is not strongly changed, while the relative flux assigned to {\feii} and broad {\hb} can be affected by the broad-line decomposition.
The narrow {\hb} luminosity has a median offset of $0.031$~dex, larger than that of {\oiii}, because narrow {\hb} lies directly under the broad {\hb} profile.
Consequently, the {\oiii}$\lambda5007$/narrow~{\hb} ratio also changes, with a median offset of $-0.036$~dex.

The derived virial quantities amplify the profile-dependent broad-{\hb} FWHM difference.
The Gaussians-based black-hole masses are shifted upward by $0.058$~dex, while the dimensionless accretion rates are shifted downward by $0.116$~dex.
This behavior follows directly from the single-epoch virial formalism:
{\mbh} depends approximately on the square of the broad-line width, and {\mdot} depends inversely on the inferred black-hole mass at fixed optical luminosity.
The offsets shown in Figure~\ref{fig:appendix_compldg} therefore support the main-text conclusion that the qualitative type~1 sequence is robust, but the derived virial quantities remain profile-dependent.

Overall, this comparison shows that the measurements least affected by the broad-{\hb} profile prescription are the continuum luminosity and {\oiii} quantities.
The most profile-dependent measurements are those tied directly to the {\hb} complex and its derived physical interpretation, including broad-{\hb} FWHM, {\rfe}, narrow {\hb} flux, {\mbh}, and {\mdot}.
These trends justify treating the broad-{\hb} profile choice as a systematic effect in the physical interpretation of low-redshift type~1 AGNs.

\section{ML-assisted refinement}
\label{app:ml}

This appendix describes the ML detector used as a supplementary check on the fitting-based broad-{\hb} type~1 classification.
ML methods have been widely applied to astronomical spectra for quasar spectral classification, line detection, and AGN classification~\citep[e.g.,][]{Busca2018,Parks2018,Falocco2022}.
The detector is not used to replace spectral decomposition or to assign the final type~1 sample.
Instead, it provides a homogeneous morphology-based score, $P_{\rm broad,H\beta}$, which quantifies how closely the {\hb}-region spectrum resembles reliable type~1 examples.
The final sample construction remains tied to the spectral-fitting measurements, while the ML score is used to remove low-confidence broad-line spectra and to flag possible missed type~1 candidates.
The ML-assisted refinement was implemented in Python using \texttt{PyTorch}~\citep{Paszke2019} and \texttt{scikit-learn}~\citep{Pedregosa2011}.
A full description of the data preparation, model training, validation, and technical implementation will be presented in a forthcoming companion paper (Li et al., in preparation).

\subsection{Input spectra and training labels}
Each spectrum was shifted to the rest frame and rebinned onto a fixed wavelength grid covering 4500--5400~{\AA}, which includes the {\hb}--{\oiii} complex and the local continuum.
Fixed-grid one-dimensional spectral representations have been widely used in deep-learning applications to astronomical spectra, including quasar spectral classification and line detection~\citep[e.g.,][]{Busca2018,Parks2018}.
To retain spectra with incomplete wavelength coverage, we adopted a two-channel input representation,
\begin{equation}
\mathbf{X} =
[\tilde{\mathbf{f}},\mathbf{m}]
\in \mathbb{R}^{2\times1024},
\end{equation}
where $\tilde{\mathbf{f}}$ is the normalized flux vector and $\mathbf{m}$ is a binary wavelength-coverage mask.
Pixels genuinely covered by the observed spectrum have $m=1$, while missing pixels have $m=0$. This representation allows the network to distinguish incomplete wavelength coverage from real spectral features.

Before training and scoring, spectra were required to pass a basic quality-control (QC) step designed to remove numerical or spectral artifacts, which assessed the distribution of normalized flux values, the flux differences between consecutive valid pixels, and the flux dispersion in the {\hb} region.
The QC criteria were used to assess input spectral quality and did not constitute physical classification criteria.

The positive training class was defined from \texttt{TYPE1\_CAND} sources, corresponding to reliable fitting-based type~1 spectra with secure broad {\hb} detections.
The comparison class was drawn from \texttt{TYPE1\_FAIL} objects, which failed the fitting-based selection.
Fitting-failed categories, including \texttt{FIT\_FAIL} objects, were excluded from the training set and were scored only after the model was trained.
Therefore, $P_{\rm broad,H\beta}$ should be interpreted as a broad-{\hb} type~1 morphology score rather than as a calibrated posterior probability for a physical type~1 classification.

\subsection{Model architecture and performance}
The detector is an ensemble of three independently initialized multi-scale one-dimensional residual neural-network models.
Each ensemble member receives the normalized flux and wavelength mask as input and returns a score $P_j$ through a sigmoid output layer.
The final broad-line score is defined as the ensemble mean,
\begin{equation}
P_{\rm broad,H\beta} = \frac{1}{3}\sum_{j=1}^{3} P_j ,
\end{equation}
and the ensemble scatter is defined as
\begin{equation}
\sigma_{\rm ens} = \left[ \frac{1}{2}\sum_{j=1}^{3} \left(P_j-P_{\rm broad,H\beta}\right)^2 \right]^{1/2}.
\end{equation}
The ensemble scatter provides a simple measure of model-to-model consistency; spectra with high $P_{\rm broad,H\beta}$ but large $\sigma_{\rm ens}$ are treated as less secure than spectra for which all ensemble members give consistent scores.

The labeled sample was split into training, validation, and independent test subsets with fractions of 70\%, 15\%, and 15\%, respectively.
The validation set was used for model selection and early stopping, while the independent test set was reserved for the final performance evaluation.
For a threshold $t$, spectra with $P_{\rm broad,H\beta}\geq t$ are classified as broad-{\hb} type~1-like objects.

We evaluate the detector using accuracy, precision, recall, $F_1$ score, and the area under the receiver operating characteristic curve (AUC).
The metrics are defined as
\begin{equation}
\left\{
\begin{array}{l}
{\rm Accuracy}= \frac{TP+TN}{TP+TN+FP+FN},\\
{\rm Precision} = \frac{TP}{TP+FP},\\
{\rm Recall} = \frac{TP}{TP+FN},\\
F_1 = \frac{2\,{\rm Precision}\,{\rm Recall}}{{\rm Precision}+{\rm Recall}}.
\end{array}
\right.
\end{equation}
where $TP$, $TN$, $FP$, and $FN$ denote true positives, true negatives, false positives, and false negatives, respectively.
Precision measures the purity of the predicted broad-{\hb} type~1-like sample, while recall measures the recovery rate of fitting-based reliable type~1 spectra.

On the independent test set, the ensemble reaches an accuracy of 96.85\%, a precision of 97.55\%, a recall of 98.48\%, an $F_1$ score of 98.01\%, and an AUC of 0.991 at the best-$F_1$ threshold, $t_{\rm best-F_1}=0.30$.
For the catalog refinement described below, we adopt a stricter threshold of $P_{\rm broad,H\beta}\geq0.8$ to define high-confidence broad-{\hb} type~1 morphology.
The stricter threshold is chosen to prioritize purity when removing low-confidence broad-line spectra and recovering missed type~1 candidates.

\begin{figure}[!t]
\centering
\includegraphics[width=\columnwidth]{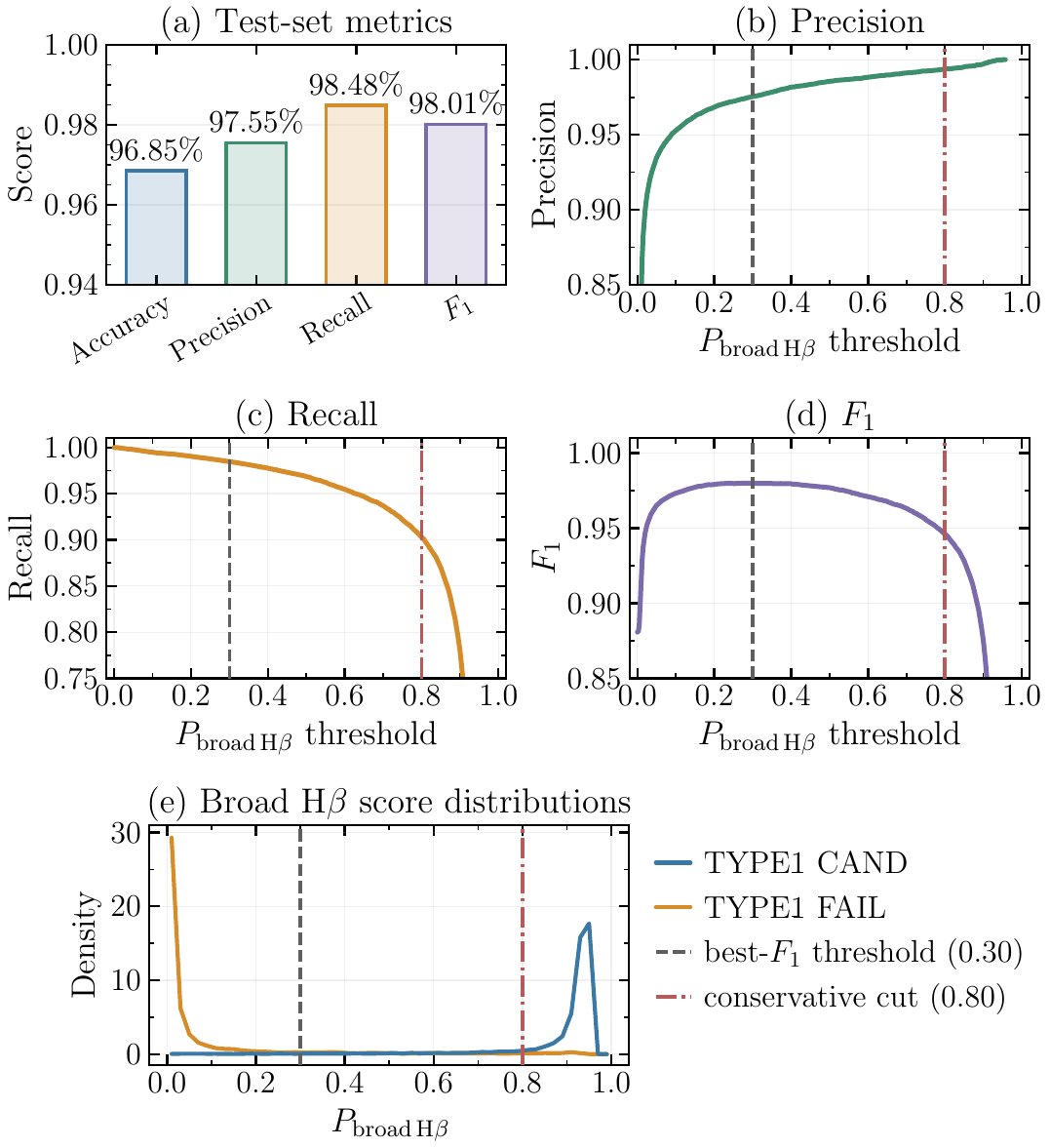}
\caption{
Test-set performance, threshold behavior, and broad-{\hb} score distributions.
Panels (a)--(d) show the independent test-set metrics and the variation of precision, recall, and $F_1$ score as a function of the threshold in $P_{\rm broad,H\beta}$.
Panel (e) shows the distributions of $P_{\rm broad,H\beta}$ after spectral quality control.
The gray dashed and red dash-dotted lines indicate the best-$F_1$ threshold ($P_{\rm broad,H\beta}=0.3$) and the conservative sample-refinement threshold ($P_{\rm broad,H\beta}=0.8$), respectively.
}
\label{fig:ml_model_performance}
\end{figure}

\subsection{Application to sample refinement}

We applied the trained detector to the catalog in two complementary steps.
First, we identified low-confidence broad-{\hb} type~1 spectra among the initially selected samples. Second, we searched the \texttt{TYPE1\_FAIL} sample for missed candidates.
In both applications, the ML output is used as a morphology-based reliability check; the final physical measurements and subclasses are still determined from spectral fitting.
This use of ML as a candidate-selection or reliability-filtering tool is similar in spirit to recent spectroscopic studies that use ML selections to identify obscured AGN populations or anomalous spectra for follow-up inspection \citep[e.g.,][]{Hviding2024,Nicolaou2026}.
For the initially selected samples, QC-passed spectra with $P_{\rm broad,H\beta}<0.8$ were removed from the high-confidence type~1 sample.
This criterion removed 12,071 sources, and the retained high-confidence sample contains 114,637 sources.

We also applied the detector to the \texttt{TYPE1\_FAIL} sources to search for missed broad-{\hb} type~1 candidates.
Since these objects failed the initial fitting-based type~1 selection, a high ML score alone is not sufficient for final inclusion.
We therefore define two recovery grades using both the mean ensemble score and the ensemble scatter:
\begin{equation}
P_{\rm broad\,H\beta}\geq0.80,\,\sigma_{\rm ens}\leq0.12,
\end{equation}
This procedure identified 588 high-confidence missed type~1 candidates in total.
These sources were flagged for refitting and visual inspection.
Only objects with reliable broad-{\hb} measurements after refitting are incorporated into the final sample construction, and their subclasses are assigned using the same broad-{\hb} FWHM criteria as for the main sample.

Overall, the ML detector provides a homogeneous supplement to the fitting-based broad~{\hb} type~1 selection.
It improves sample purity by removing low-confidence broad-{\hb} spectra from the initially selected type~1 samples and improves completeness by identifying high-confidence missed candidates.
Throughout the analysis, spectral fitting remains the basis for the final broad-line measurements and subclass definitions.

\bibliography{ref}{}
\bibliographystyle{aasjournalv7}



\end{document}